\documentclass[11pt,english,aps,pra,onecolumn,tightenlines,groupedaddress,notitlepage,floatfix,fleqn]{revtex4-1}
\pdfoutput=1
\usepackage[utf8]{inputenc}
\usepackage[T1]{fontenc}
\usepackage{amsmath}
\usepackage{amssymb}
\usepackage{amsfonts}
\usepackage{bm}
\usepackage{bbm}
\usepackage{epsfig}
\usepackage{times}
\usepackage{amsthm}
\usepackage{graphics}

\usepackage[usenames,dvipsnames]{color}
\usepackage{tikz}
\definecolor{grey}{rgb}{.35,.35,.35}
\definecolor{dblue}{rgb}{0,0.1,.6}
\definecolor{dgreen}{rgb}{0,.6,0.1}

\usepackage[colorlinks=true,citecolor=dblue,linkcolor=dblue,urlcolor=dblue]{hyperref}
\usepackage[all]{hypcap}

\newcommand{\id}{\mathbbm{1}}
\newcommand{\bra}{\langle}
\newcommand{\ket}{\rangle}
\newcommand{\bbra}{\langle\!\langle}
\newcommand{\kket}{\rangle\!\rangle}
\newcommand{\Tr}{\operatorname{Tr}}
\renewcommand{\vec}[1]{{\boldsymbol{#1}}}

\newcommand{\hGamma}{\hat{\Gamma}}
\newcommand{\hH}{\hat{H}}
\newcommand{\hL}{\hat{L}}
\newcommand{\hM}{\hat{M}}
\newcommand{\hR}{\hat{R}}
\newcommand{\hO}{\hat{O}}
\newcommand{\hA}{\hat{A}}
\newcommand{\hB}{\hat{B}}
\newcommand{\hN}{\hat{N}}
\newcommand{\hPi}{\hat{\Pi}}
\newcommand{\ha}{\hat{a}}
\newcommand{\hw}{\hat{w}}
\newcommand{\db}{{\bar{d}}}
\newcommand{\va}{\vec{a}}
\newcommand{\vb}{\vec{b}}
\newcommand{\vc}{\vec{c}}
\newcommand{\vd}{\vec{d}}
\newcommand{\vdb}{\bar{\vec{d}}}
\newcommand{\vn}{\vec{n}}
\newcommand{\dm}{{\hat{\rho}}}

\newcommand{\NN}{\mathbb{N}}
\newcommand{\RR}{\mathbb{R}}
\newcommand{\CC}{\mathbb{C}}
\newcommand{\diag}{\operatorname{diag}}
\renewcommand{\Re}{\operatorname{Re}}
\newcommand{\groupO} {\operatorname{O}}
\newcommand{\groupSp}{\operatorname{Sp}}

\newcommand{\mc}[1]{\mathcal{#1}}

\newcommand{\pdag}{{\phantom{\dag}}}

\renewcommand{\H}{\mc{H}}
\renewcommand{\L}{\mc{L}}
\newcommand{\D}{\mc{D}}
\renewcommand{\P}{\mc{P}}

\newcommand{\N}{\mc{N}}
\newcommand{\W}{\mc{W}}
\newcommand{\Wt}{\widetilde{\mc{W}}}
\newcommand{\vW} {\vec{\W}}
\newcommand{\vWt}{\widetilde{\vec{\W}}}
\newcommand{\Ns}{n}
\newcommand{\NJ}{{N_\text{J}}}

\newcommand{\mri}{\mathrm{i}}
\newcommand{\tr}{\mathrm{r}}
\newcommand{\ti}{\mathrm{i}}
\renewcommand{\ss}{\infty}

\newcommand  {\Pmatrix}[1]{\begin{pmatrix}#1\end{pmatrix}}
\newcommand  {\Psmatrix}[1]{\left(\begin{smallmatrix}#1\end{smallmatrix}\right)}

\newcommand{\Emph}[1]{\textbf{\emph{#1}}}

\newtheorem{prop}{Proposition}
\newcommand{\propHead}[1]{\bfseries \emph{#1}}

\makeatletter
\renewcommand{\p@subsection}{}
\renewcommand{\p@subsubsection}{}
\makeatother

\newcommand{\duke}  {Department of Physics, Duke University, Durham, North Carolina 27708, USA}

\begin{document}

\title{Solving quasi-free and quadratic Lindblad master equations\\ for open fermionic and bosonic systems}
\author{Thomas Barthel}
\affiliation{\duke}
\author{Yikang Zhang}
\affiliation{\duke}

\begin{abstract}
The dynamics of Markovian open quantum systems are described by Lindblad master equations. For fermionic and bosonic systems that are quasi-free, i.e., with Hamiltonians that are quadratic in the ladder operators and Lindblad operators that are linear in the ladder operators, we derive the equation of motion for the covariance matrix. This determines the evolution of Gaussian initial states and the steady states, which are also Gaussian. Using ladder super-operators (a.k.a.\ third quantization), we show how the Liouvillian can be transformed to a many-body Jordan normal form which also reveals the full many-body spectrum. Extending previous work by Prosen and Seligman, we treat fermionic and bosonic systems on equal footing with Majorana operators, shorten and complete some derivations, also address the odd-parity sector for fermions, give a criterion for the existence of bosonic steady states, cover non-diagonalizable Liouvillians also for bosons, and include quadratic systems. In extension of the quasi-free open systems, quadratic open systems comprise additional Hermitian Lindblad operators that are quadratic in the ladder operators. While Gaussian states may then evolve into non-Gaussian states, the Liouvillian can still be transformed to a useful block-triangular form, and the equations of motion for $k$-point Green's functions form a closed hierarchy. Based on this formalism, results on criticality and dissipative phase transitions in such models are discussed in a companion paper.
\end{abstract}

\date{May 2, 2022}

\maketitle

\renewcommand{\baselinestretch}{1}\normalsize
\tableofcontents
\renewcommand{\baselinestretch}{1}\normalsize

\section{Introduction}
In many cases, the time evolution of open quantum systems is to a good approximation Markovian, in the sense that the time derivative of the system's density matrix $\dm(t)$ only depends on $\dm(t)$. Typical scenarios are systems that are weakly coupled to an environment (Born-Markov and secular approximations) or systems under the influence of stochastic external fields which are generated by a Gaussian process. In Markovian systems, the dynamics is governed by a Lindblad master equation \cite{Lindblad1976-48,Gorini1976-17,Breuer2007,Rivas2012,Wolf2008-279}
\begin{equation}\label{eq:Lindblad}\textstyle
	\partial_t{\dm} = \L(\dm)=-\mri[\hH,\dm]
	 +\sum_r \Big(\hL_r^\pdag \dm \hL_r^\dag-\frac{1}{2} \{\hL_r^\dag \hL_r^\pdag,\dm\}\Big)
	 +\sum_s \Big(\hM_s^\pdag \dm \hM_s^\dag-\frac{1}{2} \{\hM_s^\dag \hM_s^\pdag,\dm\}\Big).
\end{equation}
In addition to the Hamiltonian part $-\mri[\hH,\dm]$, the Liouville super-operator $\L$ contains dissipative terms, where Lindblad operators $\hL_r$ and $\hM_s$ describe the environment couplings. 

In this work, we consider integrable systems of identical fermions or bosons, where the Hamiltonian $\hH$ and Lindblad operators $\hM_s$ are Hermitian operators that are quadratic in the ladder operators $\ha_j$ and $\ha^\dag_j$. Here, $\ha^\dag_j$ creates a particle in mode $j$, and $\ha_j$ annihilates a particle. The second considered type of Lindblad operators, $\hL_r$, is linear in the ladder operators. We call this class of open systems \emph{quadratic}, and we call the subclass that has no quadratic Lindblad operators $\hM_s$ \emph{quasi-free} \footnote{Note that the systems which we call quasi-free have often been referred to in the literature as quadratic. As discussed, we reserve the latter term for the more general class of systems that also allows for quadratic Hermitian Lindblad operators.}.
Quasi-free systems have, for example, been studied in Refs.~\cite{Prosen2008-101,Prosen2008-10,Prosen2010-07,Prosen2010-43,Eisert2010_12,Prosen2010-12,Dzhioev2011-134,Dzhioev2012-24,Koga2012-85,Yamamoto2012-370,Bardyn2013-15,Znidaric2014-112,Banchi2014-89,Budich2015-91,Venuti2016-93,Guo2017-95,Guo2018-98,Caspel2019-6,Goldstein2019-7,Teretenkov2019-22,Vernier2020-9,McDonald2022-105,Yamanaka2021_04} and quadratic systems in Refs.~\cite{Horstmann2013-87,Zunkovic2014-16,Eisler2011-06}.

\Emph{Structure and content} --
In Sec.~\ref{sec:Cov}, we introduce the covariance matrix that comprises the two-point Green's functions and derive its equation of motion for quadratic systems. We discuss fermionic and bosonic systems in parallel using Majorana operators, such that all structure matrices in the equation of motion are real. Section~\ref{sec:CovEvol} provides a first discussion on the existence of steady states, the stability of the systems, and the question whether they are relaxing.

Sec.~\ref{sec:F-Liouvillian} addresses fermionic systems. We first introduce fermionic ladder super-operators, which compose a basis for the algebra of super-operators
similar to treatments in Refs.~\cite{Schmutz1978-30,Harbola2008-465,Prosen2008-10,Dzhioev2011-134}. This formalism, which has also been called ``third quantization'' \cite{Prosen2008-10}, is then employed to express the Liouvillians of quadratic systems in a compact (third-quantized) form that conserves the number parity of the so-called super-fermions. For quasi-free systems, the Liouvillian can be transformed efficiently to a many-body Jordan normal form. Based on explicit expressions for the vacua of the correspondingly transformed annihilation super-operators, one obtains a biorthogonal operator basis in which the Liouvillian $\L$ is triangular. This gives the full many-body spectrum $\{\lambda_\vec{n}\}$ and establishes the relation between the so-called dissipative (or Liouville) gap $\Delta:=-\max_{\vn,\Re\lambda_\vn\neq 0}\Re\lambda_\vn$ and the asymptotic decay rate of the covariance-matrix evolution. Using a super-fermion number operator, the Liouvillians of quadratic systems can be brought into a block-triangular form that can be utile for assessing spectral properties like the dissipative gap \cite{Barthel2020_12}.

Section~\ref{sec:B-Liouvillian} provides an analogous discussion of the bosonic systems. The derivations are kept as parallel as possible to allow for a direct comparison to the fermionic case. Bosonic ladder super-operators are employed to obtain a third-quantized form of the Liouvillians. For the quasi-free systems, they are transformed into a many-body Jordan normal form. We then show that a many-body steady state exists if the covariance-matrix equation of motion has a physical steady-state solution. Based on the Gaussian steady state, one obtains a biorthogonal operator basis in which the Liouvillian is triangular, such that one can read off the full spectrum. Liouvillians of quadratic systems can be block-triangularized.

In Sec.~\ref{sec:Wick}, we address the question of Gaussianity. For the quasi-free case, Gaussian states evolve into Gaussian states, including the steady state, such that all observables can be computed from the covariance matrix by applying Wick's theorem. For the fermionic systems, this is also true for all elements of the constructed operators basis of (generalized) excitations. For the quadratic systems that include quadratic Lindblad operators $\hM_s$, Gaussian states generally evolve into non-Gaussian states, but the hierarchy of equations of motion for $k$-point Green's functions closes. Hence, the Green's functions can be computed with a cost that is polynomial instead of exponential in the system size.

Section~\ref{sec:B-positiveEV} revisits the questions on the existence of steady states, the stability, and the relaxation for bosonic systems. It also describes a simple example for an unstable quasi-free bosonic system with an unphysical steady-state covariance matrix and no many-body steady states.

To structure the text, central observations and results are summarized in thirteen propositions.

\Emph{Previous work} --
Let us summarize connections to the literature. For quadratic fermionic system, the covariance-matrix equation of motion was previously derived by Horstmann et al.\ \cite{Horstmann2013-87}. For the even-parity sector of quasi-free fermionic systems, the third-quantized form of the Liouvillian was previously obtained and analyzed by Prosen \cite{Prosen2008-10,Prosen2010-07}. In comparison, we also include the quadratic Lindblad terms, and we also treat the odd-parity sector. The latter is important for the determination of the many-body Jordan normal form and the full spectrum of the quasi-free systems. Finally, the steady states are given explicitly.
A third-quantized form for the Liouvillians of quasi-free bosonic systems was previously obtained and analyzed by Prosen and Seligman \cite{Prosen2010-43}. In comparison, we employ Majorana operators (algebra of position and momentum operators) such that all structure matrices are real, and we include the quadratic Lindblad terms. For the quasi-free systems, we allow for non-diagonalizable Liouvillians, and obtain a criterion for the existence of steady states as well as the stability of the systems. It is not necessary to assume that the systems are relaxing. It is sufficient that there exists a physical steady-state covariance matrix. The question whether the equations of motion for the $k$-point Green's functions form a closed hierarchy has been addressed in a rather general form by \v{Z}unkovi\v{c} \cite{Zunkovic2014-16}. We provide a short discussion for the special case of the considered quadratic systems.

\Emph{Notations} --
Operators on the Hilbert space like the Hamiltonian $\hH$ or the particle creation and annihilation operators $\ha^\dag_j$ and $\ha_j$ are indicated by a hat. Super-operators like the Liouvillian $\L$ are usually indicated by calligraphic letters, except for some ladder super-operators that are denoted by lower-case letters. We denote vectors by bold letters and employ scalar products defined as $\vec{a}\cdot\vec{b}:=\sum_j a_j b_j$.

\section{Setup and equation of motion for the covariance matrix}\label{sec:Cov}
We consider fermionic and bosonic many-body systems with an orthonormal basis of $\Ns$ single-particle states that are labeled by the index $j=1,\dots,\Ns$.
The ladder operators that annihilate or create a particle in mode $j$ are denoted by $\ha_j$ and $\ha^\dag_j=(\ha_j)^\dag$, respectively. We employ the self-adjoint Majorana operators $\hw_{j\pm}=(\hw_{j\pm})^\dag$,
\begin{equation}\label{eq:Majorna}
	\Pmatrix{\hw_{j+}\\ \hw_{j-}}:=\frac{1}{\sqrt{2}}\Pmatrix{1&1\\\mri&-\mri}\Pmatrix{\ha_j\\ \ha_j^\dag}
	\quad\Leftrightarrow\quad
	\Pmatrix{\ha_j\\ \ha_j^\dag}=\frac{1}{\sqrt{2}}\Pmatrix{1&-\mri\\1&\mri}\Pmatrix{\hw_{j+}\\ \hw_{j-}}.
\end{equation}
In the following, we discuss the evolution of single-particle correlation functions $\bra \hw_{i\mu}\hw_{j\nu}\ket=\Tr( \hw_{i\mu}\hw_{j\nu} \dm )$ that define the covariance matrix $\Gamma$ and will find the equation of motion
\begin{equation}\label{eq:EOM}\textstyle
	\partial_t \Gamma = X\Gamma+\Gamma X^T + \sum_s Z_s\Gamma Z_s^T + Y,
\end{equation}
where $X$, $Y$, and $Z_s$ are real $2\Ns\times 2\Ns$ matrices. Their definitions differ slightly for fermions and bosons; they are given below in Props.~\ref{prop:Fcov} and \ref{prop:Bcov}. $Y$ depends on matrix elements of the linear Lindblad operators $\hL_r$, $Z_s$ depends on $\hM_s$, and $X$ depends on matrix elements of $\hH$, $\hL_r$, and $\hM_s$.

\subsection{Fermionic systems}\label{sec:F-Cov}
\Emph{Setup} --
For fermions, the ladder operators $\ha_j$ and $\ha^\dag_j$ obey the canonical anti-commutation relations (CAR)
\begin{equation}\label{eq:F_a}
	\{\ha^\pdag_i,\ha^\dag_j\}\equiv \ha^\pdag_i\ha^\dag_j + \ha^\dag_j\ha^\pdag_i=\delta_{i,j}\quad\text{and}\quad
	\{\ha_i,\ha_j\}=0\quad\forall i,j.
\end{equation}
For fermions, we simplify notations by defining
\begin{equation}\label{eq:F_Majorna}
	\hw_j:=\hw_{j+} \quad\text{and}\quad \hw_{j+\Ns}:=\hw_{j-}
\end{equation}
for the Majorana operators \eqref{eq:Majorna} with $j=1,\dotsc,\Ns$. They obey the anti-commutation relations
\begin{equation}\label{eq:F_MajornaCAR}\textstyle
	\{\hw_i,\hw_j\}=\delta_{i,j} \quad\text{for}\quad i,j=1,\dotsc,2\Ns.
\end{equation}

The considered Hamiltonians
\begin{equation}\label{eq:F_H}\textstyle
	\hH=\sum_{i,j}\hw_i H_{i,j}\hw_j\quad\text{with}\quad H=H^\dag=-H^T
\end{equation}
are quadratic in the Majorana operators and, without loss of generality, the coefficient matrix $H$ can be chosen anti-symmetric due to the CAR \eqref{eq:F_MajornaCAR}, and Hermitian, as $\hH=\hH^\dag$.
The Lindblad operators $\hL_r$ are linear in the Majorana operators
\begin{equation}\label{eq:F_L-B}\textstyle
	\hL_r=\sum_{j}L_{r,j}\hw_j,\quad\text{and we define}\quad B:=\sum_{r}\vec{L}^\pdag_{r}\vec{L}^\dag_{r}\succeq 0
\end{equation}
for later use, where the vector $\vec{L}_r\in\CC^{2\Ns}$ consists of the coefficients $L_{r,j}$ such that $B_{i,j}=\sum_r L_{r,i} L^*_{r,j}$. The matrix $B$ is Hermitian and positive semidefinite. Like the Hamiltonians, the Hermitian Lindblad operators 
\begin{equation}\label{eq:F_M}\textstyle
	\hM_s=\sum_{i,j}\hw_i (M_s)_{i,j}\hw_j
\end{equation}
are quadratic in the Majorana operators and, w.l.o.g, $M_s=M_s^\dag=-M_s^T$.

\Emph{Covariance matrix} --
The covariance matrix captures all single-particle correlations and, for fermions, we define it as 
\begin{equation}\label{eq:F_G}\textstyle
	\Gamma_{i,j}
	:=\frac{\mri}{2}\bra\hw_i\hw_j-\hw_j\hw_i\ket
	\stackrel{\eqref{eq:F_MajornaCAR}}{=}\mri\bra\hw_i\hw_j\ket-\frac{\mri}{2}\delta_{i,j}
	\quad\text{such that}\quad
	\Gamma=\Gamma^*=-\Gamma^T.
\end{equation}
As $\Gamma$ is real and anti-symmetric, its eigenvalues are imaginary and come in pairs $\pm \mri\nu_k$. In particular, there exists an orthogonal transformation $U\in\groupO(2n)$ such that $UU^T=\id_{2n}$ and $\Gamma':=U\Gamma U^T=\Psmatrix{&\nu\\-\nu&}$ with the diagonal matrix $\nu$ containing all $\nu_k$. The transformation defines an alternative set of Majorana operators $\hw'_i:=\sum_j U_{i,j}\hw_j$ with CAR \eqref{eq:F_MajornaCAR} and covariance matrix $\Gamma'$ such that $\nu_k=\mri\bra \hw'_{k+}\hw'_{k-}\ket$. As each occupation number operator $\ha_j^\dag\ha^\pdag_j$ has eigenvalues $0$ and $1$, the relation \eqref{eq:Majorna} implies that the operators $\mri\hw_{j+}\hw_{j-}={1}/{2}-\ha_j^\dag\ha^\pdag_j$ and the operators $\mri\hw'_{k+}\hw'_{k-}$ have eigenvalues $\pm 1/2$. It follows that the spectrum $\{\nu_k\}$ of physical covariance matrices is in the interval $[-{1}/{2},{1}/{2}]$. Furthermore, all covariance matrix elements obey 
\begin{equation}\textstyle
	|\Gamma_{i,j}|\leq \|U^T\Gamma' U\|=\|\nu\oplus(-\nu)\|\leq \frac{1}{2}.
\end{equation}

\Emph{Equation of motion} --
To obtain the equation of motion for $\Gamma$, we go to the Heisenberg picture, i.e., consider the operator time evolution
\begin{subequations}
\begin{alignat}{4}\label{eq:LindbladAdj}
	&\partial_t \hO = \L^\dag(\hO)&&=\mri[\hH,\hO] + \D_L^\dag(\hO) + \D_M^\dag(\hO)\\
	\label{eq:LindbladAdj-DL}
	&\text{with}\quad \D_L^\dag(\hO)&&\textstyle=\sum_r \big(\hL_r^\dag \hO \hL_r^\pdag-\frac{1}{2} \{\hL_r^\dag \hL_r^\pdag,\hO\}\big)\\
	\label{eq:LindbladAdj-DM}
	&\text{and}\quad \D_M^\dag(\hO)&&\textstyle=\sum_s \big(\hM_s^\dag \dm \hM_s^\pdag-\frac{1}{2} \{\hM_s^\dag \hM_s^\pdag,\hO\}\big)
	  = -\frac{1}{2}\sum_s[\hM_s,[\hM_s,\hO]]
\end{alignat}
\end{subequations}
Using
\begin{equation}\label{eq:F-comm-ww-w}\textstyle
	[\hw_i\hw_j,\hw_m]=\delta_{j,m}\hw_i - \delta_{i,m}\hw_j,\quad\text{and with}\quad \hGamma_{m,n}:=\mri\hw_m\hw_n-\frac{\mri}{2}\delta_{m,n}
\end{equation}
being a $2\Ns\times 2\Ns$ matrix of operators such that $\Gamma=\bra \hGamma\ket$, one finds that
\begin{subequations}
\begin{alignat}{4}\nonumber
	&[\hH,\hw_m\hw_n]&&=[\hH,\hw_m]\hw_n + \hw_m[\hH,\hw_n]\\ \nonumber
	&&&\textstyle=\sum_i H_{i,m}\hw_i\hw_n - \sum_j H_{m,j}\hw_j\hw_n + \sum_i H_{i,n}\hw_m\hw_i - \sum_j H_{n,j}\hw_m\hw_j\\ \textstyle
	\label{eq:Fcov-H}
	&\text{and hence} &&[\hH,\hGamma_{m,n}]=-2\big(H\hGamma+\hGamma H^T\big)_{m,n}.
\end{alignat}
Here, $\big(H\hGamma\big)_{m,n}\equiv\sum_j H_{m,j}\hGamma_{j,n}$.
For the dissipator \eqref{eq:LindbladAdj-DL} of the linear Lindblad operators \eqref{eq:F_L-B}, we have
\begin{alignat}{4}\nonumber
	&\D_L^\dag(\hw_m\hw_n)&&\textstyle=\frac{1}{2}\sum_r\sum_{i,j}L_{r,j}L^*_{r,i}\left(\hw_i[\hw_m\hw_n,\hw_j]-[\hw_m\hw_n,\hw_i]\hw_j\right)\\\nonumber
	&&&\textstyle= \frac{1}{2}\sum_{i}\left( B_{n,i}\hw_i\hw_m - B_{m,i}\hw_i\hw_n\right)
	              -\frac{1}{2}\sum_{j}\left( B_{j,n}\hw_m\hw_j - B_{j,m}\hw_n\hw_j\right)\\
	\label{eq:Fcov-DL}
	&\text{and hence} &&\textstyle\D_L^\dag(\hGamma_{m,n})=-\big(\frac{B+B^*}{2}\,\hGamma+\hGamma\, \frac{B+B^*}{2}-\frac{B-B^*}{2\mri}\big)_{m,n}.
\end{alignat}
For the dissipator \eqref{eq:LindbladAdj-DM} of the quadratic Hermitian Lindblad operators \eqref{eq:F_M}, we can simply apply the result \eqref{eq:Fcov-H} for the Hamiltonian term twice to obtain
\begin{equation}\label{eq:Fcov-DM}\textstyle
	\D_M^\dag(\hGamma_{m,n})=-\frac{1}{2}\sum_s[\hM_s,[\hM_s,\hGamma_{m,n}]]
	 = 2\sum_s\big(2M_s \hGamma M_s - M_s^2\hGamma-\hGamma M_s^2 \big)_{m,n}.
\end{equation}
\end{subequations}
Putting together the results in Eqs.~\eqref{eq:Fcov-H}-\eqref{eq:Fcov-DM} and taking the expectation value in $\partial_t\Gamma=\bra \L^\dag(\hGamma) \ket$, we obtain the equation of motion \eqref{eq:EOM}.
\begin{prop}[\propHead{Evolution of fermionic covariance matrices}]\label{prop:Fcov}
The covariance matrix \eqref{eq:F_G} of Markovian fermionic systems with a quadratic Hamiltonian \eqref{eq:F_H}, linear Lindblad operators \eqref{eq:F_L-B}, and quadratic Hermitian Lindblad operators \eqref{eq:F_M} evolves according to the equation of motion \eqref{eq:EOM} with the contained real $2\Ns\times 2\Ns$ matrices given by
\begin{equation}\label{eq:F-EOM}\textstyle
	X=-2\mri H-B_\tr-2\sum_s M_s^2,\quad
	Y=B_\ti,\quad\text{and}\quad
	Z_s=2\mri M_s.
\end{equation}
\end{prop}
Here, the symmetric real part $B_\tr$ and the anti-symmetric imaginary part $B_\ti$ of the matrix $B$ from Eq.~\eqref{eq:F_L-B} are defined as
\begin{equation}\label{eq:Br-Bi}
	B_\tr:=(B+B^*)/2=B_\tr^T\quad\text{and}\quad
	B_\ti:=(B-B^*)/2\mri=-B_\ti^T.
\end{equation}

\subsection{Bosonic systems}\label{sec:B-Cov}
\Emph{Setup} --
For bosons, the ladder operators $\ha_j$ and $\ha^\dag_j$ obey the canonical commutation relations (CCR)
\begin{equation}\label{eq:B_a}
	[\ha^\pdag_i,\ha^\dag_j]\equiv \ha^\pdag_i\ha^\dag_j - \ha^\dag_j\ha^\pdag_i=\delta_{i,j}\quad\text{and}\quad
	[\ha_i,\ha_j]=0\quad\forall i,j.
\end{equation}
Note that the Majorana operators defined in Eq.~\eqref{eq:Majorna} are basically position operators ($\hw_{j+}=\hat{x}_j$) and momentum operators ($\hw_{j-}=-\hat{p}_j$) for the modes $j=1.\dotsc,\Ns$. They obey the commutation relations
\begin{equation}\label{eq:B_MajornaCCR}\textstyle
	[\hw_{i\mu},\hw_{j\nu}]=-\mri\mu\delta_{i,j}\delta_{\mu,\bar{\nu}},\quad\text{where}\quad
	\mu,\nu=\pm 1\quad\text{and}\quad \bar{\nu}:=-\nu.
\end{equation}

The considered Hamiltonians are quadratic in the ladder operators such that
\begin{equation}\label{eq:B_H}\textstyle
	\hH=\sum_{i\mu,j\nu}\hw_{i\mu} H_{i\mu,j\nu}\hw_{j\nu}\quad\text{with}\quad H=H^T=H^*.
\end{equation}
That, without loss of generality, the coefficient matrix $H$ can be chosen to be real and symmetric follows from the Hermiticity of $\hH$ and the commutation relations \eqref{eq:B_MajornaCCR}:
\begin{equation}\label{eq:tau}\textstyle
	\hH\stackrel{\eqref{eq:B_MajornaCCR}}{=}\frac{1}{2}\sum_{i\mu,j\nu}\hw_{i\mu}(H_{i\mu,j\nu}+H_{j\nu,i\mu}) \hw_{j\nu}-\frac{1}{2}\Tr(\tau H)
	\ \ \text{with}\ \ \tau:=\sigma_y\otimes \id_{\Ns}=\Psmatrix{0&-\mri\id_{\Ns}\\ \mri\id_{\Ns}&0}.
\end{equation}
Hence, we can choose $H=H^T$. For the matrix notations in the above equation and in the following, we use the mode-ordering $(1+,2+,\dotsc,\Ns+,1-,2-,\dotsc,\Ns-)$, $\sigma_y$ is the Pauli matrix $\Psmatrix{0&-\mri\\ \mri&0}$, and $\id_{\Ns}$ is the $\Ns\times \Ns$ identity matrix such that $\Tr(\tau H)=0$. From $\hH=\hH^\dag$ follows that $H$ can be chosen real.
The Lindblad operators $\hL_r$ are linear in the Majorana operators
\begin{equation}\label{eq:B_L-B}\textstyle
	\hL_r=\sum_{j\nu}L_{r,j\nu}\hw_{j\nu}\quad\text{and we define}\quad B:=\sum_{r}\vec{L}^\pdag_{r}\vec{L}^\dag_{r}\succeq 0
\end{equation}
as for fermions. The vector $\vec{L}_r\in\CC^{2\Ns}$ consists of the coefficients $L_{r,j\nu}$ and the matrix $B$ is Hermitian and positive semidefinite. Like the Hamiltonians, the Hermitian Lindblad operators 
\begin{equation}\label{eq:B_M}\textstyle
	\hM_s=\sum_{i\mu,j\nu}\hw_{i\mu} (M_s)_{i\mu,j\nu}\hw_{j\nu}
\end{equation}
are quadratic in the Majorana operators and $M_s=M_s^T=M_s^*$.

\Emph{Covariance matrix} --
We define the bosonic covariance matrix as 
\begin{equation}\label{eq:B_G}\textstyle
	\Gamma_{i\mu,j\nu}
	:=\frac{1}{2}\bra\hw_{i\mu}\hw_{j\nu}+\hw_{j\nu}\hw_{i\mu}\ket
	\stackrel{\eqref{eq:B_MajornaCCR}}{=}\bra\hw_{i\mu}\hw_{j\nu}\ket+\frac{\mri}{2}\mu\delta_{i,j}\delta_{\mu,\bar{\nu}}
	\quad\text{such that}\quad
	\Gamma=\Gamma^*=\Gamma^T.
\end{equation}
The matrix $\Gamma$ is not only real and symmetric, but also positive definite: For any vector $\vec{v}\in \CC^{2n}$, $\vec{v}^*\cdot\Gamma\vec{v}=\bra \hat{V}^\dag\hat{V}+\hat{V}\hat{V}^\dag\ket/2\geq 0$ is the expectation value of a positive semidefinite operator, where $\hat{V}=\sum_{j\nu}v_{j\nu}\hw_{j\nu}$. Hence, $\Gamma$ is positive semidefinite.
Let $G_{i\mu,j\nu}:=\bra\hw_{i\mu}\hw_{j\nu}\ket/2$ such that $G=G^\dag$ and $\Gamma=G+G^T$. If $\Gamma$ had not full rank, there would exist an eigenvector $\vec{g}$ such that $\Gamma\vec{g}=\vec{0}$. As $G$ and $G^T$ are both positive semidefinite, this would also imply $G\vec{g}=\vec{0}$ and $G^T\vec{g}=\vec{0}$. This is a contradiction because, according to the commutation relations \eqref{eq:B_MajornaCCR}, $G=\tau/2+G^T$ such that $G-G^T=\tau/2$ has full-rank, i.e., has an empty kernel. Thus, $\Gamma$ is positive definite.

Then, according to Williamson's theorem \cite{Williamson1936-58,Simon1999-40}, there exists a real symplectic transformation $U\in\groupSp(n)$ such that $U\tau U^T=\tau$ and $\Gamma':=U\Gamma U^T=\Psmatrix{\nu&\\&\nu}$ with the diagonal matrix $\nu$ containing the symplectic eigenvalues $\nu_k$. The transformation defines an alternative set of Majorana operators $\hw'_{k\kappa}:=\sum_{j\nu} U_{k\kappa,j\nu}\hw_{j\nu}$ with CCR \eqref{eq:B_MajornaCCR} and covariance matrix $\Gamma'$ such that $\nu_k=\bra \hw'_{k+}\hw'_{k+}\ket=\bra \hw'_{k-}\hw'_{k-}\ket$. As each occupation number operator $\ha_j^\dag\ha^\pdag_j$ has eigenvalues $0,1,2,\dotsc$, the relation \eqref{eq:Majorna} implies that the operators $\frac{1}{2}(\hw_{j+}\hw_{j+}+\hw_{j-}\hw_{j-})=\ha_j^\dag\ha^\pdag_j+{1}/{2}$ and the operators $\frac{1}{2}(\hw'_{k+}\hw'_{k+}+\hw'_{k-}\hw'_{k-})$ have eigenvalues $1/2,3/2,\dotsc$. It follows that the symplectic spectrum $\{\nu_k\}$ of physical covariance matrices is in the interval $[{1}/{2},\infty)$.

\Emph{Equation of motion} --
As for fermions in Sec.~\ref{sec:F-Cov}, one obtains the equation of motion for $\Gamma$ by considering the operator time-evolution according to the adjoint Lindblad master equation \eqref{eq:LindbladAdj}. Using
\begin{equation}\label{eq:B-comm-ww-w}\textstyle
	[\hw_{i\alpha}\hw_{j\beta},\hw_{m\mu}]
	=\mri\mu(\delta_{j,m}\delta_{\beta,\bar{\mu}}\hw_{i\alpha} + \delta_{i,m}\delta_{\alpha,\bar{\mu}}\hw_{j\beta}),
	\ \ \text{and}\ \ \hGamma_{m\mu,n\nu}:=\hw_{m\mu}\hw_{n\nu} +\frac{\mri}{2}\mu\delta_{m,n}\delta_{\mu,\bar{\nu}}
\end{equation}
being a $2\Ns\times 2\Ns$ matrix of operators such that $\Gamma=\bra \hGamma\ket$, one finds that
\begin{subequations}
\begin{alignat}{4}\nonumber
	&[\hH,\hw_{m\mu}\hw_{n\nu}]&&=[\hH,\hw_{m\mu}]\hw_{n\nu} + \hw_{m\mu}[\hH,\hw_{n\nu}]\\ \nonumber
	&&&\textstyle=-\mu\left(\sum_{i\alpha} H_{i\alpha,m\bar{\mu}}\hw_{i\alpha}\hw_{n\nu}
	                       +\sum_{j\beta}  H_{m\bar{\mu},j\beta}\hw_{j\beta}\hw_{n\nu} \right)\\  \nonumber
	&&&\textstyle\phantom{=}-\nu\left(\sum_{i\alpha} H_{i\alpha,n\bar{\nu}}\hw_{m\mu}\hw_{i\alpha}
	                       +\sum_{j\beta}  H_{n\bar{\nu},j\beta}\hw_{m\mu}\hw_{j\beta} \right)\\ \textstyle
	\label{eq:Bcov-H}
	&\text{and hence} &&[\hH,\hGamma_{m\mu,n\nu}]=-2\big(\tau H\,\hGamma+\hGamma\, (\tau H)^T\big)_{m,n}.
\end{alignat}
For the dissipator \eqref{eq:LindbladAdj-DL} of the linear Lindblad operators \eqref{eq:B_L-B}, we have
\begin{alignat}{4}\nonumber
	&\D_L^\dag(\hw_{m\mu}\hw_{n\nu})&&\textstyle
	=\frac{1}{2}\sum_r\sum_{i\alpha,j\beta}L_{r,j\beta}L^*_{r,i\alpha}\left(\hw_{i\alpha}[\hw_{m\mu}\hw_{n\nu},\hw_{j\beta}]-[\hw_{m\mu}\hw_{n\nu},\hw_{i\alpha}]\hw_{j\beta}\right)\\\nonumber
	&&&\textstyle= -\frac{\mri}{2}\sum_{i\alpha}\left( \nu B_{n\bar{\nu},i\alpha}\hw_{i\alpha}\hw_{m\mu}
	                                                 + \mu B_{m\bar{\mu},i\alpha}\hw_{i\alpha}\hw_{n\nu}\right)\\\nonumber
	&&&\textstyle\phantom{=} +\frac{\mri}{2}\sum_{j\beta} \left( \nu B_{j\beta, n\bar{\nu}}\hw_{m\mu}\hw_{j\beta}
	                                                 + \mu B_{j\beta, m\bar{\mu}}\hw_{n\nu}\hw_{j\beta}\right)\\
	\label{eq:Bcov-DL}
	&\text{and hence} &&\textstyle
	\D_L^\dag(\hGamma_{m,n})=\big(\tau\frac{B-B^*}{2}\,\hGamma+\hGamma\,(\tau\frac{B-B^*}{2})^T+\tau\frac{B+B^*}{2}\tau\big)_{m,n}.
\end{alignat}
For the dissipator \eqref{eq:LindbladAdj-DM} of the quadratic Hermitian Lindblad operators \eqref{eq:B_M}, we can apply the result \eqref{eq:Bcov-H} for the Hamiltonian term twice to obtain
\begin{align}\nonumber
	\D_M^\dag(\hGamma_{m\mu,n\nu})
	 &\textstyle=-\frac{1}{2}\sum_s[\hM_s,[\hM_s,\hGamma_{m\mu,n\nu}]]\\ \label{eq:Bcov-DM}
	 &\textstyle= -2\sum_s\big(2\tau M_s \hGamma (\tau M_s)^T + (\tau M_s)^2\hGamma+\hGamma ((\tau M_s)^2)^T \big)_{m\mu,n\nu}.
\end{align}
\end{subequations}
Putting together the results in Eqs.~\eqref{eq:Bcov-H}-\eqref{eq:Bcov-DM} and taking the expectation value in $\partial_t\Gamma=\bra \L^\dag(\hGamma) \ket$, we obtain the equation of motion \eqref{eq:EOM}.
\begin{prop}[\propHead{Evolution of bosonic covariance matrices}]\label{prop:Bcov}
The covariance matrix \eqref{eq:B_G} of Markovian bosonic systems with quadratic Hamiltonians \eqref{eq:B_H}, linear Lindblad operators \eqref{eq:B_L-B}, and quadratic Hermitian Lindblad operators \eqref{eq:B_M} evolves according to the equation of motion \eqref{eq:EOM} with the contained real $2\Ns\times 2\Ns$ matrices given by
\begin{equation}\label{eq:B-EOM}\textstyle
	X=-2\mri\tau H+\mri\tau B_\ti-2\sum_s (\tau M_s)^2,\quad
	Y=\tau B_\tr\tau,\quad\text{and}\quad
	Z_s=2\mri \tau M_s.
\end{equation}
\end{prop}
The matrix $\tau$ was defined in Eq.~\eqref{eq:tau}. $B_\tr$ and $B_\ti$, as defined in Eq.~\eqref{eq:Br-Bi}, are the real and imaginary parts of the matrix $B$ that characterizes the linear Lindblad operators \eqref{eq:B_L-B}.

\subsection{Covariance-matrix evolution, single-particle spectrum, and steady states}\label{sec:CovEvol}
In the previous sections, we saw that the covariance matrices $\Gamma$ [Eqs.~\eqref{eq:F_G} and \eqref{eq:B_G}] for quadratic fermionic and bosonic systems evolve according to the linear first-order differential equation \eqref{eq:EOM}. Hence, steady-state covariance matrices $\Gamma_\ss$ obey the linear equation
\begin{equation}\label{eq:Gss}\textstyle
	X\Gamma_\ss+\Gamma_\ss X^T + \sum_s Z_s\Gamma_\ss Z_s^T = -Y.
\end{equation}
For quasi-free systems, which have no quadratic Lindblad operators $\hM_s$ and hence $Z_s=0$, it assumes the form of a Sylvester equation or, more specifically, a continuous Lyapunov equation \cite{Parks1992-9,Simoncini2016-58}. These play important roles in control theory and stability analysis \cite{Sastry1999,Khalil2002}.

To discuss the solution to the equation of motion \eqref{eq:EOM} and the spectrum of its generator, it is convenient to write it in the vectorized form
\begin{equation}\label{eq:EOMvec}\textstyle
	\partial_t \vec{\Gamma} = K\vec{\Gamma} + \vec{Y} \quad\text{with generator}\quad
	K:=X\otimes\id+\id\otimes X + \sum_s Z_s\otimes Z_s=K_+\oplus K_-.
\end{equation}
Here, the real $2\Ns\times 2\Ns$ matrices $\Gamma$ and $Y$ have been vectorized into $4\Ns^2$ dimensional vectors $\vec{\Gamma}$ and $\vec{Y}$ such that $(\vec{\Gamma})_{(i-1)2\Ns+j}=\Gamma_{i,j}$ for $i,j=1,\dots,2\Ns$ in accordance with the tensor product. Note that for bosonic systems, only the symmetric subspace is physically relevant as $\Gamma^T=\Gamma$. For fermions, $\Gamma^T=-\Gamma$ such that only the anti-symmetric subspace matters. Let isometries $P_+$ and $P_-$ denote the projectors for these subspaces. As the generator $K$ commutes with the swap operator, it assumes the block-diagonal form $K=K_+\oplus K_-$, where $K_\pm:=P_\pm K P^\dag_\pm$. For the following, let $\vec{\Gamma}\mapsto P_\pm \vec{\Gamma}$ and $\vec{Y}\mapsto P_\pm \vec{Y}$ denote the projections to the (anti-)symmetric subspace.

For invertible $K_\pm$, the solution for an evolution starting from covariance vector $\vec{\Gamma}(0)$ at time $t=0$ is
\begin{equation}\label{eq:EOMvecSol}\textstyle
	\vec{\Gamma}(t) = e^{K_\pm t}\left(\vec{\Gamma}(0) + K_\pm^{-1}\vec{Y} \right) - K_\pm^{-1}\vec{Y}.
\end{equation}
The relaxation to the steady-state covariance matrix $\vec{\Gamma}_\ss=-K_\pm^{-1}\vec{Y}$ is governed by the generator $K_\pm$, and the steady state is unique if $K_\pm$ is invertible.

For the system to be \emph{stable}, i.e., to not allow for a divergence of covariance matrix elements, all eigenvalues $\kappa_m$ of $K_\pm$ must have non-positive real parts.
For the system to be \emph{relaxing}, i.e., converging to a unique steady-state covariance matrix, independent of the initial condition $\Gamma(0)$, all eigenvalues $\kappa_m$ must have strictly negative real parts \footnote{In stability theory, such matrices are called \emph{stable} or \emph{Hurwitz} \cite{Sastry1999,Khalil2002}.}. $K_\pm$ is in general not Hermitian and can be non-diagonalizable. One hence needs to consider its Jordan normal form and generalized eigenvectors.

Fermionic systems are always stable as, according to the definition \eqref{eq:F_G}, all elements of the covariance matrix are bounded by $-1/2\leq \Gamma_{i,j}\leq 1/2$. This is reflected in the fact that all $K_-$ eigenvalues have non-positive real parts: In particular, let $\vec{k}$ be a normalized (anti-symmetric) ordinary eigenvector of $K$ with eigenvalue $\kappa$. Then, using $X$, $Z_s$, and $Y$ as given in Eq.~\eqref{eq:F-EOM}
\begin{align}\textstyle\nonumber
	\kappa =    \vec{k}^\dag K \vec{k}
	 = &   -2\mri \vec{k}^\dag(H\otimes\id+\id\otimes H)\vec{k}
	            - \vec{k}^\dag(B_\tr\otimes\id+\id\otimes B_\tr)\vec{k}\\ 
	   &\textstyle\label{eq:EOM-Keval}
	       -2\sum_s\vec{k}^\dag(M_s^2\otimes\id+\id\otimes M_s^2 + 2 M_s\otimes M_s)\vec{k}.
\end{align}
Due to the Hermiticity of $H$, the first term yields an imaginary number. As $B_\tr$ is Hermitian and positive semidefinite, the second term gives a non-positive real number. The matrix of the third term can be written as  $(M_s\otimes\id+\id\otimes M_s)^2$ which is clearly positive semidefinite such that the third term also contributes a non-positive real number, and $\Re\kappa\leq 0$.

For a finite number $n$ of single-particle states (modes), fermionic systems always have at least one steady state and, hence, there exists a physical covariance matrix $\Gamma_\ss$ that solves Eq.~\eqref{eq:Gss}. This is because the density operators for a finite-dimensional system form a compact convex set. With the time evolution operator $e^{\L t}$ being a continuous map, Brouwer's fixed-point theorem \cite{Brouwer1911-71,Kakutani1941-8} asserts the existence of a steady state.

For bosonic systems, the situation is different. The elements of the covariance matrix are unbounded, and the generator $K_+$ can in fact have positive eigenvalues, i.e., there exist unstable systems. Physically, this corresponds to situations where the environment indefinitely pumps energy (and particles) into the system. Furthermore, the set of density operators is not compact, and the system may hence not have any steady state. A simple example with Lindblad operator $\hL_i=\ha^\dag_i$ is discussed in Sec.~\ref{sec:B-positiveEV}.

\section{Liouvillians for fermionic systems and their Jordan normal form}\label{sec:F-Liouvillian}
In the following, we will introduce fermionic ladder super-operators (Sec.~\ref{sec:F-superLadder}) that form a basis for the algebra of super-operators. They are then employed to express the Liouvillians of quadratic fermion systems in a compact (a.k.a.\ third-quantized) form that conserves the number parity of the so-called super-fermions (Sec.~\ref{sec:F-3rdQuantization}). The Liouvillians of quasi-free systems are quadratic in the ladder super-operators, which implies that we can easily determine eigenvalues and (generalized) eigenstates and that Wick's theorem is applicable. We show how to transform the Liouvillians of quasi-free systems to a many-body Jordan normal form (Sec.~\ref{sec:F-Quasi-free}) and obtain an explicit expression for the Gaussian steady state (Sec.~\ref{sec:F-Quasi-free-ss}). One can then construct a biorthogonal operator basis such that the Liouvillian assumes a triangular form and the spectrum can be read off (Sec.~\ref{sec:F-Quasi-free-Spectrum}). Exploiting that the number of super-fermions is non-decreasing under the action of the Liouvillian, Liouvillians of quadratic systems can be brought into a useful block-triangular form (Sec.~\ref{sec:F-blockTriangular}).

\subsection{Ladder super-operators}\label{sec:F-superLadder}
In Sec.~\ref{sec:F-Cov}, we wrote the Hamiltonian $\hH$ and Lindblad operators $\hL_r$ and $\hM_s$ in terms of the Majorana operators \eqref{eq:F_Majorna}. To express the action of the Liouvillian on an operator $\hR$, it is convenient to define the super-operators
\begin{subequations}\label{eq:F-c}
\begin{align}
	c_j(\hR):=&\textstyle\frac{1}{\sqrt{2}}\big(\hw_j\hR-\hPi\hR\hPi\hw_j\big)\\ \label{eq:F-c+}
	\Rightarrow\ 
	c^\dag_j(\hR)=&\textstyle\frac{1}{\sqrt{2}}\big(\hw_j\hR+\hPi\hR\hPi\hw_j\big),
\end{align}
\end{subequations}
where the particle-number parity operator $\hPi$ is defined as
\begin{equation}\label{eq:parityOp}\textstyle
	\hPi=(-1)^{\hN}\quad\text{with}\quad \hN=\sum_{j=1}^\Ns\ha^\dag_j\ha^\pdag_j
\end{equation}
being the particle-number operator.
The adjoint in Eq.~\eqref{eq:F-c+} is defined through the Hilbert-Schmidt inner product
\begin{equation}\label{eq:HS-innerProd}
	\bbra \hA|\hB\kket:=\Tr(\hA^\dag\hB)
\end{equation}
such that
\begin{equation*}\textstyle
	\bbra \hA|c_j(\hB)\kket
	=\frac{1}{\sqrt{2}}\Tr\left(\hA^\dag\big(\hw_j\hB-\hPi\hB\hPi\hw_j\big)\right)
	=\frac{1}{\sqrt{2}}\Tr\left(\big(\hA^\dag\hw_j+\hw_j\hPi\hA^\dag\hPi\big)\hB\right)
	=\bbra c^\dag_j(\hA)|\hB\kket
\end{equation*}
for any two operators $\hA$ and $\hB$ on the Hilbert space, where we have used that $\hw_j\hPi=-\hw_j\hPi$.
Straightforward calculations show that the ladder super-operators \eqref{eq:F-c} obey the canonical anti-commutation relations
\begin{equation}\label{eq:F-c-CAR}\textstyle
	\{c_i,c_j^\dag\}=\delta_{i,j},\quad
	\{c_i,c_j\}=0,\quad\text{and}\quad
	\{c^\dag_i,c^\dag_j\}=0\quad\text{for}\quad i,j=1,\dotsc,2\Ns.
\end{equation}
For example,
\begin{align*}\textstyle
	2\{c_i,c_j^\dag\}(\hR)
	=&\,\,\phantom{+}\hw_i\hw_j\hR + \hw_i\hPi\hR\hPi\hw_j - \hPi\hw_j\hR\hPi\hw_i - \hR\hPi\hw_j\hPi\hw_i\\
	 &          +\hw_j\hw_i\hR - \hw_j\hPi\hR\hPi\hw_i + \hPi\hw_i\hR\hPi\hw_j - \hR\hPi\hw_i\hPi\hw_j
	=2\delta_{i,j}\hR,
\end{align*}
because terms 1 and 5 as well as 4 and 8 give $\delta_{i,j}\hR$, terms 2 and 7 cancel, and terms 3 and 6 cancel. The other two anti-commutation rules in Eq.~\eqref{eq:F-c-CAR} follow similarly.

The definition \eqref{eq:F-c} of the fermionic ladder super-operators is analogous to corresponding definitions in Refs.\ \cite{Schmutz1978-30,Harbola2008-465,Prosen2008-10,Dzhioev2011-134}. The approach is known as \emph{super-fermion formalism} or \emph{third quantization} \footnote{Like the term \emph{second quantization} \cite{Negele1988}, the name \emph{third quantization} \cite{Prosen2008-10,Prosen2010-43} should be taken with a grain of salt. Neither of the two involves an actual quantization. The former is a formalism to describe closed quantum systems of identical particles with ladder operators and a Fock basis for the many-body Hilbert space $\mathcal{H}$; the latter is a corresponding formalism for open systems. Instead of thinking in terms of super-operators, one can also interpret the ladder super-operators of third quantization as usual (second quantization) ladder operators on the enlarged Hilbert space $\mathcal{H}\otimes\mathcal{H}$, corresponding to the vectorization of operators that maps from the Banach space $\mathcal{B}(\mathcal{H})$ to $\mathcal{H}\otimes\mathcal{H}$.}.

\subsection{Liouvillian in third quantization}\label{sec:F-3rdQuantization}
To express the Liouvillian \eqref{eq:Lindblad} in terms of the ladder super-operators \eqref{eq:F-c}, we first introduce the additional self-adjoint super-operators
\begin{subequations}\label{eq:F-W}
\begin{align}
	\W_j(\hR) :=&\hw_j\hR=\textstyle\frac{1}{\sqrt{2}}\big(c^\dag_j+c_j\big)(\hR)\quad \text{and}\\
	\Wt_j(\hR):=&\hR\hw_j=\textstyle\frac{1}{\sqrt{2}}\big(c^\dag_j-c_j\big)\P(\hR)=\frac{1}{\sqrt{2}}\P\big(c_j-c^\dag_j\big)(\hR),
\end{align}
\end{subequations}
where $\P(\hR):=\hPi\hR\hPi$ conjugates $\hR$ with the parity operator \eqref{eq:parityOp}, and $\P^2(\hR)=\hR$.
The elementary Hamiltonian and linear Lindblad terms for the Liouvillian \eqref{eq:Lindblad} can now be expressed in the form
\begin{subequations}\label{eq:F-Wterms}
\begin{align}
	[\hw_i\hw_j,\hR]\label{eq:F-W-comm}
	&\textstyle= \big(\W_i\W_j-\Wt_j\Wt_i\big)(\hR)\quad\text{and}\\\textstyle
	\hw_i\hR\hw_j-\frac{1}{2}\{\hw_j\hw_i,\hR\}\label{eq:F-W-Lindblad}
	&\textstyle= \big(\W_i\Wt_j-\frac{1}{2}(\W_j\W_i+\Wt_i\Wt_j)\big)(\hR)
\end{align}
\end{subequations}
Defining the vectors $\vc:=(c_1,\dotsc,c_{2\Ns})^T$, $\vc^\dag:=(c^\dag_1,\dotsc,c^\dag_{2\Ns})^T$, $\vW:=(\W_1,\dotsc,\W_{2\Ns})^T$, and $\vWt:=(\Wt_1,\dotsc,\Wt_{2\Ns})^T$, using Eq.~\eqref{eq:F-W-comm} and the relations
\begin{equation}\label{eq:F-W-vs-c}
	\Pmatrix{\vW\\ \vWt}=\frac{1}{\sqrt{2}}\Pmatrix{\id_{2\Ns}&\id_{2\Ns}\\ \P\id_{2\Ns}&-\P\id_{2\Ns}} \Pmatrix{\vc\\ \vc^\dag},\quad
	\Pmatrix{\vW\\ \vWt}^T=\frac{1}{\sqrt{2}}\Pmatrix{\vc\\ \vc^\dag}^T \Pmatrix{\id_{2\Ns}&-\P\id_{2\Ns}\\ \id_{2\Ns}&\P\id_{2\Ns}},
\end{equation}
the commutator with the Hamiltonian \eqref{eq:F_H} assumes the form
\begin{equation}\label{eq:F-L3rd-H}
	[\hH,\hR]
	= \Pmatrix{\vW\\ \vWt}    \cdot \Pmatrix{H&\\&H}\Pmatrix{\vW\\ \vWt}(\hR)
	= \Pmatrix{\vc\\ \vc^\dag}\cdot \Pmatrix{&H\\H&}\Pmatrix{\vc\\ \vc^\dag}(\hR)
	=2\vc^\dag\cdot H\vc\,(\hR)=2\vc\cdot H\vc^\dag\,(\hR).
\end{equation}
Here, it has been used that $H^T=-H$. Also, recall that scalar products are denoted by $\vec{a}\cdot\vec{b}=\sum_j a_j b_j=\vec{a}^T\vec{b}$. For the quadratic Hermitian Lindblad operators \eqref{eq:F_M} immediately follows
\begin{equation}\textstyle\label{eq:F-L3rd-M}
	\D_M(\hR)=-\frac{1}{2}\sum_s[\hM_s,[\hM_s,\hR]]
	= - 2\sum_s\big(\vc^\dag\cdot M_s\vc\big)^2\,(\hR).
\end{equation}
The contribution of the linear Lindblad operators \eqref{eq:F_L-B} is a bit more complex and depends on the particle-number parity. Using Eq.~\eqref{eq:F-W-Lindblad}, $B^T=B^*$, and the relations \eqref{eq:F-W-vs-c}, we find
\begin{align}\textstyle\nonumber
	\D_L(\hR)
	&= \sum_r \Big(\hL_r^\pdag \hR \hL_r^\dag-\frac{1}{2} \{\hL_r^\dag \hL_r^\pdag,\hR\}\Big)
	 \textstyle\stackrel{\eqref{eq:F-W-Lindblad}}{=}
	   \left(\vW\cdot B\vWt-\frac{1}{2}\big(\vW\cdot B^T\vW+\vWt\cdot B\vWt\big)\right) (\hR)\\\nonumber
	&= \frac{1}{2}\Pmatrix{\vW\\ \vWt} \cdot \Pmatrix{-B^* & B\\B^*&-B}\Pmatrix{\vW\\ \vWt}(\hR)
	 \stackrel{\eqref{eq:F-W-vs-c}}{=}
	   \Pmatrix{\vc\\ \vc^\dag} \cdot \Pmatrix{\mri B_\ti\P_+ & -B_\tr\P_+\\-B_\tr\P_- & \mri B_\ti\P_-}\Pmatrix{\vc\\ \vc^\dag}(\hR)\\
	&\textstyle\label{eq:F-L3rd-L}
	 = \left(-\vc^\dag\cdot B_\tr\vc\, \P_+ + \vc^\dag\cdot \mri B_\ti\vc^\dag\, \P_+
	         -\vc\cdot B_\tr\vc^\dag\, \P_- + \vc\cdot \mri B_\ti\vc\, \P_-\right)(\hR).
\end{align}
Here, $B_\tr$ and $B_\ti$ are the real and imaginary parts of the matrix $B$ as defined in Eq.~\eqref{eq:Br-Bi}, 
\begin{equation}\textstyle
	\P_\pm(\hR):=\frac{1\pm\P}{2}(\hR) = \frac{1}{2}\big(\hR\pm \hPi\hR\hPi\big)
\end{equation}
are the projections onto the even and odd parity sectors, and we have used $\P_\pm\vc^{(\dag)}=\vc^{(\dag)}\P_\mp$.
Combining the results in Eqs.~\eqref{eq:F-L3rd-H}-\eqref{eq:F-L3rd-L}, we have the Liouvillian in terms of the ladder super-operators \eqref{eq:F-c}.
\begin{prop}[\propHead{Quadratic fermionic Liouvillians in third quantization}]\label{prop:F-3rdQuantization}
The Liouvillian $\L$ in Eq.~\eqref{eq:Lindblad} for Markovian fermionic systems with quadratic Hamiltonian \eqref{eq:F_H}, linear Lindblad operators \eqref{eq:F_L-B}, and quadratic Hermitian Lindblad operators \eqref{eq:F_M} conserves the particle-number parity, $\L\P=\P\L$. Using the ladder super-operators \eqref{eq:F-c}, it can hence be written as a direct sum $\L=\L_+\oplus\L_-$ with the Liouvillian
\begin{subequations}\label{eq:F-L3rd}
\begin{equation}\label{eq:F-Lp}\textstyle
	\L_+=\vc^\dag\cdot X_0\vc + \vc^\dag\cdot \mri B_\ti\vc^\dag - 2\sum_s (\vc^\dag\cdot M_s\vc)^2
\end{equation}
for the even-parity sector and
\begin{equation}\label{eq:F-Lm}\textstyle
	\L_-=\vc\cdot X_0\vc^\dag + \vc\cdot \mri B_\ti\vc - 2\sum_s (\vc\cdot M_s\vc^\dag)^2
\end{equation}
\end{subequations}
for the odd parity sector, where $X_0:=-2\mri H-B_\tr$.
\end{prop}

The normal-ordered form of $\L_+$ and anti-normal-ordered form of $\L_-$ read
\begin{subequations}\begin{align}
	\L_+ &\label{eq:F-Lp-no}\textstyle
	= \underbrace{\textstyle\vc^\dag\cdot \big(X_0-2\sum_s M_s^2\big)\vc + \vc^\dag\cdot \mri B_\ti\vc^\dag}_{
	        =\vc^\dag\cdot X\vc + \vc^\dag\cdot iY\vc^\dag}
	  + 2\sum_s \sum_{i,j,k,\ell}(M_s)_{i,j}(M_s)_{k,\ell}c_i^\dag c_k^\dag c_j^\pdag c_\ell^\pdag,\\
	\L_- &\label{eq:F-Lm-no}\textstyle
	= \underbrace{\textstyle\vc\cdot \big(X_0-2\sum_s M_s^2\big)\vc^\dag + \vc\cdot \mri B_\ti\vc}_{
	        =\vc\cdot X\vc^\dag + \vc\cdot \mri Y\vc}
	  + 2\sum_s \sum_{i,j,k,\ell}(M_s)_{i,j}(M_s)_{k,\ell}c_i^\pdag c_k^\pdag c_j^\dag c_\ell^\dag,
\end{align}
\end{subequations}
where we have the same matrices $X$ and $Y$ as in Eq.~\eqref{eq:F-EOM}. This establishes the connection to the covariance-matrix equation of motion, discussed in Sec.~\ref{sec:F-Cov}.

\subsection{Many-body Jordan normal form for quasi-free systems}\label{sec:F-Quasi-free}
For quasi-free systems, i.e., systems comprising only linear Lindblad operators $\hL_r$ but \emph{no} quadratic ones ($\hM_s$), the Liouvillians \eqref{eq:F-L3rd} for the even and odd-parity sectors assume the forms
\begin{equation}\label{eq:F-Lfree}
	\L_+ = \frac{1}{2}\Pmatrix{\vc\\ \vc^\dag}\cdot \Pmatrix{&-X_0^T\\X_0& 2\mri Y}\Pmatrix{\vc\\ \vc^\dag}+\frac{\Tr X_0}{2},
	\quad
	\L_- = \frac{1}{2}\Pmatrix{\vc\\ \vc^\dag}\cdot \Pmatrix{2\mri Y&X_0\\-X_0^T&}\Pmatrix{\vc\\ \vc^\dag}+\frac{\Tr X_0}{2},
\end{equation}
where $X_0=-2\mri H-B_\tr$ and $Y=B_\ti$ in accordance with Eq.~\eqref{eq:F-EOM}.

Let us bring $\L_+$ into a many-body Jordan normal form through a canonical (Bogoliubov) transformation that leads to new ladder super-operators
\begin{equation}\label{eq:F-dV}
	\Pmatrix{\vd\\ \vd'}:=V\Pmatrix{\vc\\ \vc^\dag}\quad\text{such that}\quad
	(V^{-1})^T \Pmatrix{&-X_0^T\\X_0& 2\mri Y} V^{-1} = \Pmatrix{&-\xi^T\\\xi &},
\end{equation}
where $\xi$ is a $2\Ns\times 2\Ns$ matrix in Jordan normal form with $\NJ$ Jordan blocks with eigenvalues $\xi_k$ and dimensions $D_k$.
Notice that we write $\vd'$ instead of $\vd^\dag$ as, in general, we will have $d_{k,\ell}'\neq (d_{k,\ell})^\dag$.
For the new ladder operators to obey the canonical anti-commutation relations, we need that
\begin{equation}\label{eq:F-d-CAR}
	J_-:= \Pmatrix{ & \id_{2\Ns}\\ \id_{2\Ns} & }
	= \Pmatrix{\{\vd,\vd^T\} & \{\vd,\vd'^T\} \\ \{\vd',\vd^T\} & \{\vd',\vd'^T\} } 
	= V \Pmatrix{\{\vc,\vc^T\} & \{\vc,\vc^{\dag T}\} \\ \{\vc^\dag,\vc^T\} & \{\vc^\dag,\vc^{\dag T}\} } V^T 
	= V J_- V^T,
\end{equation}
i.e., the transformation $V$ needs to leave $J_-=\sigma_x\otimes \id_{2\Ns}$ invariant. 
Such a transformation does indeed exist \cite{Prosen2010-07}: Let $S$ be the similarity transformation that brings $X_0$ to Jordan normal form $\xi$,
\begin{equation}
	S^{-1}X_0 S=\xi,\quad\text{and make the ansatz}\quad
	V=\underbrace{\Pmatrix{S^{-1}& \\&S^T}}_{=:V_S}
	  \underbrace{\Pmatrix{\id_{2\Ns}& -2\mri\Gamma\\&\id_{2\Ns}}}_{=:V_\Gamma}
\end{equation}
with $\Gamma=-\Gamma^T\in\RR^{2\Ns\times 2\Ns}$. This choice for $V$ obeys Eq.~\eqref{eq:F-d-CAR} as
\begin{equation*}
	J_-\,V_S^T J_-=\Pmatrix{S& \\&(S^{-1})^T}=V_S^{-1}\quad\text{and}\quad
	J_-\,V_\Gamma^T J_-=\Pmatrix{\id_{2\Ns}& 2\mri\Gamma\\&\id_{2\Ns}}=V_\Gamma^{-1}.
\end{equation*}
And, for the Liouvillian $\L_+$, we obtain
\begin{equation}\label{eq:F-Lp-d}
	\L_+-\frac{\Tr X_0}{2}
	=\frac{1}{2}\Pmatrix{\vd\\ \vd'}\cdot (V^{-1})^T \Pmatrix{&-X_0^T\\X_0& 2\mri Y} V^{-1}\Pmatrix{\vd\\ \vd'}
	=\frac{1}{2}\Pmatrix{\vd\\ \vd'}\cdot \Pmatrix{&-\xi^T\\\xi & Q} \Pmatrix{\vd\\ \vd'},
\end{equation}
where $Q=2\mri S^{-1}(X_0\Gamma+\Gamma X_0^T+Y)(S^{-1})^T$ is zero if we choose $\Gamma$ as a steady-state covariance matrix $\Gamma_\ss$, i.e. a solution of Eq.~\eqref{eq:Gss} with $M_s=Z_s=0$. As discussed in Sec.~\ref{sec:CovEvol}, finite fermionic systems have at least one steady state. With $Q=0$ and $\Tr X_0=\Tr\xi=\sum_k D_k \xi_k$, we arrive at $\L_+=\vd'\cdot\xi\vd$.
One can proceed analogously for $\L_-$ by simply switching the roles of $c_i$ and $c_i^\dag$ as suggested by Eq.~\eqref{eq:F-Lfree}.
\begin{prop}[\propHead{Jordan normal form for quasi-free fermionic Liouvillians}]\label{prop:F-L-Jordan}
The Liouvillian $\L$ in Eq.~\eqref{eq:Lindblad} for Markovian fermionic systems with quadratic Hamiltonian \eqref{eq:F_H} and linear Lindblad operators \eqref{eq:F_L-B} can be transformed to a many-body Jordan normal form. Starting from the third-quantized form \eqref{eq:F-Lfree}, using a steady-state covariance matrix $\Gamma_\ss$ that solves the continuous Lyapunov equation \cite{Parks1992-9,Simoncini2016-58,Sastry1999,Khalil2002}
\begin{subequations}\label{eq:F-L-Jordan}
\begin{equation}\label{eq:F-Gss-free}\textstyle
	X_0\Gamma_\ss+\Gamma_\ss X_0^T = -Y\quad\text{with}\quad
	X_0=-2\mri H-B_\tr\quad\text{and}\quad
	Y=B_\ti,
\end{equation}
and using the transformation of $X_0$ to Jordan normal form $S^{-1}X_0 S=\xi$, the Liouvillian for the even-parity sector becomes
\begin{equation}\label{eq:F-Lp-Jordan}
	\L_+=\sum^\NJ_{k=1}\left(\xi_k\sum_{\ell=1}^{D_k}d'_{k,\ell}d_{k,\ell} + \sum_{\ell=1}^{D_k-1}d'_{k,\ell}d_{k,\ell+1}\right)
	\quad\!\!\text{with}\quad\!\!
	\Pmatrix{\vd\\ \vd'}=\Pmatrix{S^{-1}& \\&S^T}\Pmatrix{\id& -2\mri\Gamma_\ss\\&\id}\Pmatrix{\vc\\ \vc^\dag}.
\end{equation}
Here, the $k$th Jordan block of $X_0$ has eigenvalue $\xi_k$ (with $\Re\xi_k\leq 0$) and dimension $D_k$, $d'_{k,\ell}$ creates a super-fermion (excitation) that corresponds to the generalized $\xi_k$ eigenvector of rank $\ell$, and the ladder super-operators obey the anti-commutation relations
\begin{equation}\label{eq:F-d-CAR2}\textstyle
	\{d_{k,\ell},d'_{k',\ell'}\}=\delta_{k,k'}\delta_{\ell,\ell'},\quad
	\{d_{k,\ell},d_{k',\ell'}\}=0,\quad\text{and}\quad
	\{d'_{k,\ell},d'_{k',\ell'}\}=0\quad \forall k,k',\ell,\ell'.
\end{equation}
Switching the roles of $c_i$ and $c_i^\dag$, the Liouvillian for the odd-parity sector assumes the same form
\begin{equation}\label{eq:F-Lm-Jordan}
	\L_-=\sum^\NJ_{k=1}\left(\xi_k\sum_{\ell=1}^{D_k}\db'_{k,\ell}\db_{k,\ell} + \sum_{\ell=1}^{D_k-1}\db'_{k,\ell}\db_{k,\ell+1}\right)
	\quad\!\!\text{for}\quad\!\!
	\Pmatrix{\vdb\\ \vdb'}=\Pmatrix{S^{-1}& \\&S^T}\Pmatrix{\id& -2\mri\Gamma_\ss\\&\id}\Pmatrix{\vc^\dag\\ \vc}.
\end{equation}
\end{subequations}
The ladder super-operators $\db_{k,\ell}$ and $\db'_{k,\ell}$ also obey the canonical anti-commutation relations \eqref{eq:F-d-CAR2}.
\end{prop}

That all $X_0$ eigenvalues $\xi_k$ have non-positive real-parts follows similar to the discussion of Eq.~\eqref{eq:EOM-Keval}: Let $\vec{\xi}$ be a normalized (ordinary) eigenvector of $X_0$ with eigenvalue $\xi$. Then $\xi=\vec{\xi}^\dag X_0 \vec{\xi}=-2\mri\vec{\xi}^\dag H\vec{\xi}-\vec{\xi}^\dag B_r\vec{\xi}$. Due to the Hermiticity of $H$, the first term contributes an imaginary number. As $B_\tr$ is positive-semidefinite, the second term contributes a non-positive real number such that $\Re\xi\leq 0$.

\subsection{Steady states for quasi-free systems}\label{sec:F-Quasi-free-ss}
Operators $\hR$ in the sector with odd parity (difference), $\hPi\hR\hPi=-\hR$, are all traceless. Steady states must hence have support in the even-parity sector.
Given a steady-state covariance matrix $\Gamma_\ss$ that solves the continuous Lyapunov equation \eqref{eq:F-Gss-free}, we will use the the many-body Jordan normal form \eqref{eq:F-Lp-Jordan} of the Liouvillian \eqref{eq:F-Lp} for the even-parity sector to establish that the Gaussian state $\dm(\Gamma_\ss)$ with covariance matrix $\Gamma_\ss$ is a steady state. If the system has eigenvalues $\xi_k=0$, further steady states exist, including traceless contributions from the odd-parity sector.

For the following, we employ a \emph{Dirac notation} with super-kets $|\hB\kket:=\hB$ and super-bras $\bbra\hA|$, where $\hA$ and $\hB$ are operators on the Hilbert space $\H$, such that $\bbra \hA|\hB\kket=\Tr(\hA^\dag\hB)$ in accordance with the Hilbert-Schmidt inner product \eqref{eq:HS-innerProd}.

We will use that, according to Wick's theorem, \emph{Gaussian states} are fully characterized by their covariance matrix \cite{Wick1950-80,Danielewicz1984-152,Negele1988}. In particular, let the covariance matrix $\Gamma$ be diagonalized by the orthogonal transformation $U$ in the sense that $U\Gamma U^T=\Psmatrix{&\nu\\-\nu&}$ with $\nu=\diag(\nu_1,\dotsc,\nu_n)$, and consider the transformed set of Majorana operators $\hw'_i=\sum_j U_{i,j}\hw_j$ as discussed in Sec.~\ref{sec:F-Cov}. Using that a Gaussian state $\dm$ is the exponential of an operator that is quadratic in the ladder operators, $\Tr(\dm)=1$, and $\nu_k=\mri\Tr(\dm  \hw'_{k+}\hw'_{k-})$, the Gaussian state $\dm$ with covariance matrix $\Gamma$ can be written explicitly in the form
\begin{equation}\label{eq:F-GaussianState}\textstyle
	\dm=\dm(\Gamma)=\prod_k\left({1}/{2}+\nu_k\right)\,\left(\frac{{1}/{2}-\nu_k}{{1}/{2}+\nu_k}\right)^{\mri\hw'_{k-}\hw'_{k+}+1/2}
	\quad\text{with}\quad \nu_k\in[-1/2,1/2]
\end{equation}
and eigenvalues $\prod_k \left({1}/{2}+\nu_k\right)\,\left(\frac{{1}/{2}-\nu_k}{{1}/{2}+\nu_k}\right)^{n_k}$ for $n_k\in\{0,1\}$.

\begin{prop}[\propHead{Steady states and vacua for quasi-free fermionic Liouvillians}]\label{prop:F-ss}
Finite quasi-free fermionic systems have at least one steady state. Let $\Gamma_\ss$ be a steady-state covariance matrix. The corresponding Gaussian state $\dm(\Gamma_\ss)$ is then a steady state for the Liouvillian $\L$ [Eq.~\eqref{eq:F-L3rd}], has even parity (difference), and is simultaneously the right vacuum of the annihilation super-operators $d_{k,\ell}$ defined in Eq.~\eqref{eq:F-Lp-Jordan}, i.e.,
\begin{subequations}
\begin{equation}\label{eq:F-Quasi-free-vac-d}
	d_{k,\ell}|\vec{0}\kket_d=0\ \ \forall\ k,\ell,\quad
	\L|\vec{0}\kket_d=0,
	\quad\text{and}\quad
	\bbra \id|\vec{0}\kket_d=\Tr \dm(\Gamma_\ss)=1 \quad\text{for}\quad
	|\vec{0}\kket_d:=\dm(\Gamma_\ss).
\end{equation}
If the matrix $X_0$ in Prop.~\ref{prop:F-L-Jordan} has zero eigenvalues $\xi_k=0$, then applying an even number of the corresponding creation super-operators $d'_{k,1}$ to $|\vec{0}\kket_d$ and forming linear combinations yields further (Gaussian) steady states.
The right vacuum of the annihilation super-operators $\db_{k,\ell}$ for the odd-parity sector as defined in Eq.~\eqref{eq:F-Lm-Jordan} is $\hPi\dm(\Gamma_\ss)$, i.e.,
\begin{equation}\label{eq:F-Quasi-free-vac-db}
	\db_{k,\ell}|\vec{0}\kket_\db=0\ \ \forall\ k,\ell
	\quad\text{and}\quad
	\bbra \hPi|\vec{0}\kket_\db=\Tr \dm(\Gamma_\ss)=1 \quad\text{for}\quad
	|\vec{0}\kket_\db:=\hPi\dm(\Gamma_\ss).
\end{equation}
If the matrix $X_0$ in Prop.~\ref{prop:F-L-Jordan} has zero eigenvalues $\xi_k=0$, then applying an odd number of the corresponding super-operators $\db'_{k,1}$ to $|\vec{0}\kket_\db$ yields traceless operators. Forming linear combinations with the steady states described above yields further steady states.
Lastly, $\bbra\id|$ and $\bbra\hPi|$ are the left vacua for the creation super-operators of the even and odd-parity sectors, respectively,
\begin{equation}\label{eq:F-Quasi-free-left-vac}
	\bbra\id|d'_{k,\ell}=0\quad\text{and}\quad
	\bbra\hPi|\db'_{k,\ell}=0 \quad\forall\ k,\ell.
\end{equation}
\end{subequations}
\end{prop}

As discussed in Sec.~\ref{sec:CovEvol}, quasi-free fermionic systems are always stable and, for finite systems sizes $n$, they have at least one steady-state covariance matrix $\Gamma_\ss$ which can be determined by solving the continuous Lyapunov equation \eqref{eq:F-Gss-free}. To see that the corresponding Gaussian state $|\vec{0}\kket_d\equiv \dm(\Gamma_\ss)=:\dm_\ss$ is the vacuum for all annihilation super-operators $d_{k,\ell}$, consider the matrix elements 
\begin{equation}\label{eq:F-d-matrixElements}
	\bbra\id|\mc{A}\, d_{k,\ell}|\vec{0}\kket_d=\Tr [\mc{A}\, d_{k,\ell}(\dm_\ss)],
\end{equation}
where $\mc{A}$ is an arbitrary product of ladder super-operators like $d_{5,1}d'_{2,1}d_{5,3}d'_{3,1}d'_{3,2}$. As the latter form a super-operator basis, we can conclude that $d_{k,\ell}|\vec{0}\kket_d=0$ if the matrix elements \eqref{eq:F-d-matrixElements} are zero for all $\mc{A}$. The matrix element can only be nonzero if the number of factors in $\mc{A}$ is odd, because Gaussian states \eqref{eq:F-GaussianState} have even parity difference, $\hPi\dm_\ss\hPi=\dm_\ss$.

For fermionic Gaussian states $\dm$ and $\{\mc{D}_i\}$ being linear combinations of ladder (super-)operators, Wick's theorem \cite{Wick1950-80,Danielewicz1984-152,Negele1988} establishes that
\begin{equation}\label{eq:F-Wick}\textstyle
	\bra\mc{D}_1\dotsb \mc{D}_{2m}\ket=\sum_{\vec{i}\in\operatorname{Pair}_m} (-1)^{\sigma(\vec{i})}
	\bra \mc{D}_{i_1}\mc{D}_{i_2}\ket\dotsb \bra \mc{D}_{i_{2m-1}}\mc{D}_{i_{2m}}\ket,
\end{equation}
where $\operatorname{Pair}_m$ is the set of all partitions $\vec{i}$ of numbers $\{1,\dotsc,2m\}$ into pairs without respect of order, i.e., $\vec{i}=\{(i_1,i_2),\ldots,(i_{2m-1},i_{2m})\}$ with $i_{2k-1}<i_{2k}$ $\forall\ k$. There are $(2m-1)!!=(2m-1)(2m-3)\dotsb 1$ of such pairings, and $\sigma(\vec{i})$ denotes the number of crossings in the pair contraction $\vec{i}$.

The single-particle correlations (two-point Green's functions) needed for the evaluation of the matrix elements \eqref{eq:F-d-matrixElements} are
\begin{equation}\label{eq:F-d-correlMatrix}
	\bbra\id| \Pmatrix{\vd\,\vd^T & \vd\,\vd'^T \\ \vd'\,\vd^T & \vd'\,\vd'^T } |\vec{0}\kket_d
	= \, V \,\bbra\id|\Pmatrix{\vc\,\vc^T & \vc\,\vc^{\dag T} \\ \vc^\dag\,\vc^T & \vc^\dag\,\vc^{\dag T} }|\vec{0}\kket_d \,V^T,
\end{equation}
where $V$ denotes the Bogoliubov transformation in Eqs.~\eqref{eq:F-dV} and \eqref{eq:F-Lp-Jordan}. Let us now determine the four $2n\times 2n$ blocks on the right-hand side. The property $c_j(\id)=0$ $\forall j$ implies $\bbra \id|\vc^\dag=\vec{0}$ such that the lower two blocks are both zero, i.e., $\bbra\id|\vc^\dag\,\vc^T|\vec{0}\kket_d=0$ and $\bbra\id|\vc^\dag\,\vc^{\dag T}|\vec{0}\kket_d=0$. Due to the anti-commutation relations \eqref{eq:F-c-CAR}, the upper-right block is $\bbra\id|\vc\,\vc^{\dag T}|\vec{0}\kket_d=\id_{2n} - \bbra\id|\vc^\dag\,\vc^T|\vec{0}\kket_d=\id_{2n}$. For the upper-left block, note that
\begin{equation*}\textstyle
	c_i c_j(\dm_\ss)\stackrel{\eqref{eq:F-c}}{=}\frac{1}{2}\left(\hw_i\hw_j\dm_\ss-\hw_i\dm_\ss\hw_j+\hw_j\dm_\ss\hw_i-\dm_\ss\hw_j\hw_i\right),
\end{equation*}
and, hence, $\Gamma_\ss=\frac{\mri}{2}\bbra \id|\vc \vc^T|\vec{0}\kket_d$ according to the definition \eqref{eq:F_G}. Combining these results, the single-particle correlation matrix \eqref{eq:F-d-correlMatrix} evaluates to
\begin{equation}\label{eq:F-d-correlMatrix2}
	\bbra\id| \Pmatrix{\vd\,\vd^T & \vd\,\vd'^T \\ \vd'\,\vd^T & \vd'\,\vd'^T } |\vec{0}\kket_d
	= V \,\Pmatrix{-2\mri\Gamma_\ss & \id_{2n} \\ 0 & 0 } \,V^T 
	\stackrel{\eqref{eq:F-Lp-Jordan}}{=} \Pmatrix{0 & \id_{2n} \\ 0 & 0 }.
\end{equation}
Now, when expanding matrix elements \eqref{eq:F-d-matrixElements} through Wick's theorem \eqref{eq:F-Wick} into a product of single-particle correlation functions, there is always a factor $\bra \D_i\, d_{k,\ell}\ket\equiv \bbra\id|\D_i\, d_{k,\ell}|\vec{0}\kket_d$. This factor is zero because, according to Eq.~\eqref{eq:F-d-correlMatrix2} both $\bbra\id|\vd'\,\vd^T|\vec{0}\kket_d=0$ and $\bbra\id|\vd\,\vd^T|\vec{0}\kket_d=0$. Thus, the matrix elements \eqref{eq:F-d-matrixElements} are always zero, which establishes that $d_{k,\ell}|\vec{0}\kket_d=0$ $\forall\ k,\ell$ and, according to Eq.~\eqref{eq:F-Lp-Jordan}, $\L|\vec{0}\kket_d=\L_+|\vec{0}\kket_d=0$.
If $X_0$ in Prop.~\ref{prop:F-L-Jordan} has zero eigenvalues $\xi_k=0$, then further steady states are obtained by acting with an even number of the corresponding creation super-operators on $|\vec{0}\kket_d$ and forming linear combinations. One needs an even number to remain in the even-parity sector. In particular, $\L_+ d'_{k,1}|\vec{0}\kket_d=d'_{k,1}\L_+|\vec{0}\kket_d=0$ if $\xi_k=0$.

Analogously, it follows that $\bbra\id|$ is the left vacuum of $\vd'$ as the lower two blocks in Eq.~\eqref{eq:F-d-correlMatrix2} vanish. But this can also be seen more directly as
\begin{equation}\label{F:left-d-vacuum}
	\bbra \id|\vc^\dag\stackrel{\eqref{eq:F-c}}{=}\vec{0}\quad\Rightarrow\quad
	\bbra\id|\vd'\stackrel{\eqref{eq:F-Lp-Jordan}}{=}S^T\bbra\id|\vc^\dag=\vec{0}.
\end{equation}

To prove Eq.~\eqref{eq:F-Quasi-free-vac-db}, first note that $\bbra \hPi|\vec{0}\kket_\db=\Tr( \hPi^2\dm_\ss)=1$ as $\hPi^2=\id$. Second, note that $\hPi c_j(\hPi\hR)=-c_j^\dag(\hR)$ according to the definition \eqref{eq:F-c} and, hence, $\hPi \db_{k,\ell}(\hPi\hR)=-d_{k,\ell}$ as well as $\hPi \db'_{k,\ell}(\hPi\hR)=-d'_{k,\ell}$ according to Eqs.~\eqref{eq:F-Lp-Jordan} and \eqref{eq:F-Lm-Jordan}.
It then follows that the operator $|\vec{0}\kket_\db$ is the vacuum of all annihilation super-operators $\db_{k,\ell}$, because the matrix elements 
\begin{equation*}
	\bbra\hPi|\bar{\mc{A}}\, \db_{k,\ell}|\vec{0}\kket_\db
	\stackrel{\eqref{eq:F-Quasi-free-vac-db}}{=}\Tr \big[\hPi\bar{\mc{A}}\, \db_{k,\ell}(\hPi\dm_\ss)\big]
	= - \Tr \big[\hPi\bar{\mc{A}}\big(\hPi\, d_{k,\ell}(\dm_\ss)\big)\big]
	= - \Tr \big[\mc{A} \,d_{k,\ell}(\dm_\ss)\big]=0
\end{equation*}
can be expressed in terms of the previously discussed matrix elements \eqref{eq:F-d-matrixElements} and are, hence, zero
for any product $\bar{\mc{A}}$ of ladder super-operators like $\db_{5,1}\db'_{2,1}\db_{5,3}\db'_{3,1}\db'_{3,2}$. Here, $\mc{A}=\hPi\bar{\mc{A}}\hPi$ is the corresponding product like $-d_{5,1}d'_{2,1}d_{5,3}d'_{3,1}d'_{3,2}$.
If the matrix $X_0$ in Prop.~\ref{prop:F-L-Jordan} has zero eigenvalues $\xi_k=0$, then applying an odd number of the corresponding super-operators $\db'_{k,1}$ to $|\vec{0}\kket_\db$ yields traceless operators with odd parity difference. These can be combined linearly with even-parity steady states to obtain further steady states.

Lastly, $\bbra\hPi|$ is the left vacuum of $\vdb'$ as 
\begin{equation*}
	\bbra \hPi|\vc\stackrel{\eqref{eq:F-c}}{=}\vec{0}\quad\Rightarrow\quad
	\bbra\hPi|\vdb'\stackrel{\eqref{eq:F-Lm-Jordan}}{=}S^T\bbra\hPi|\vc=\vec{0}.
\end{equation*}

\subsection{Triangular form and spectrum for quasi-free systems}\label{sec:F-Quasi-free-Spectrum}
\begin{prop}[\propHead{Triangular form and spectrum for quasi-free fermionic Liouvillians}]\label{prop:F-spectrum}
With the ladder super-operators for the Jordan normal form \eqref{eq:F-L-Jordan} of quasi-free fermionic Liouvillians and their corresponding left and right vacua described in Prop.~\ref{prop:F-ss}, the right and left operator bases
\begin{subequations}
\begin{equation}\label{eq:F-basisEven}\textstyle
	|\vn\kket:={\overrightarrow{\prod}}_{k,\ell}(d'_{k,\ell})^{n_{k,\ell}}|\vec{0}\kket_d,\quad
	\bbra\vn|:= \bbra\id|{\overleftarrow{\prod}}_{k,\ell}(d_{k,\ell})^{n_{k,\ell}}\quad\text{with even}\ \
	\sum_{k,\ell}n_{k,\ell}
\end{equation}
for the even-parity sector and 
\begin{equation}\label{eq:F-basisOdd}\textstyle
	|\vn\kket:={\overrightarrow{\prod}}_{k,\ell}(\db'_{k,\ell})^{n_{k,\ell}}|\vec{0}\kket_\db,\quad
	\bbra\vn|:= \bbra\hPi|{\overleftarrow{\prod}}_{k,\ell}(\db_{k,\ell})^{n_{k,\ell}}\quad\text{with odd}\ \
	\sum_{k,\ell}n_{k,\ell}
\end{equation}
for the odd-parity sector form a biorthogonal basis, i.e., $\bbra\vn|\vn'\kket=\delta_{\vn,\vn'}$ $\forall$ $\vn,\vn'$.
Here, the arrows above the product symbols indicate the ordering in the products of the ladder super-operators, and
$\vn:=(n_{1,1},\dotsc,n_{1,D_1},n_{2,1},\dotsc,n_{\NJ,D_\NJ})$ is the vector of (super-fermion) occupation numbers with $n_{k,\ell}\in\{0,1\}$. If we order the bases $\{|\vn\kket\}$ and $\{\bbra\vn|\}$ according to increasing $I_\vn:=\sum^\NJ_{k=1}\sum_{\ell=1}^{D_k} \ell\, n_{k,\ell}$, then the matrix representation
\begin{equation}\label{eq:F-L-upperTriag}
	\L_{\vn,\vn'}:=\bbra\vn|\L|\vn'\kket
\end{equation}
of the Liouvillian is upper-triangular with its diagonal containing the (generalized) eigenvalues
\begin{equation}\label{eq:F-spectrum}\textstyle
	\{\lambda_\vn = \sum^\NJ_{k=1}\sum_{\ell=1}^{D_k}\xi_k n_{k,\ell}\,|\,n_{k,\ell}=0,1\}.
\end{equation}
The system is relaxing with unique steady state $|\vec{0}\kket_d$ if $\Re\xi_k<0$ for all generalized eigenvalues of $X_0$ in Eq.~\eqref{eq:F-Gss-free}. The dissipative gap is then $\Delta\equiv-\max_{\vn\neq \vec{0}}\Re\lambda_\vn=-\max_k\Re\xi_k$, corresponding to a single super-fermion excitation in the odd-parity sector.
\end{subequations}
\end{prop}

The operators \eqref{eq:F-basisEven}, denoted in the Dirac notation for the Hilbert-Schmidt inner product \eqref{eq:HS-innerProd}, have even parity (difference). That is, $\P|\vn\kket=|\vn\kket$ and $\bbra\vn|\P=\bbra\vn|$ with $\P$ as defined below Eq.~\eqref{eq:F-W}, because the right vacuum $|\vec{0}\kket_d$ and the identity $\bbra\id|$ have even parity, and we apply an even number of ladder super-operators, each flipping the parity. Similarly, $|\vec{0}\kket_\db$ and the parity operator $\bbra\hPi|$ have even parity and, hence, the operators \eqref{eq:F-basisOdd} have odd parity. Of course, the inner product $\bbra\vn|\vn'\kket$ is zero if one operator has odd and the other has even parity. The biorthogonality of two even-parity operators follows from the anti-commutation relations \eqref{eq:F-d-CAR2}, $\vd|\vec{0}\kket_d=\vec{0}$, $\bbra \id|\vd'=\vec{0}$, and $\bbra \id|\vec{0}\kket_d=1$ (cf.\ Prop.~\ref{prop:F-ss});
\begin{align*}
	\bbra\vn|\vn'\kket&=\textstyle \bbra\id|
	\left({\overleftarrow{\prod}}_{k,\ell}(d_{k,\ell})^{n_{k,\ell}}\right)
	\left({\overrightarrow{\prod}}_{k,\ell}(d'_{k,\ell})^{n'_{k,\ell}}\right)
	|\vec{0}\kket_d\\
	&=\textstyle
	\prod_{k,\ell}\delta_{n_{k,\ell},n'_{k,\ell}}\bbra\id|\vec{0}\kket_d\pm
	\bbra\id|
	\left({\overrightarrow{\prod}}_{k,\ell}(d'_{k,\ell})^{n'_{k,\ell}}\right)
	\left({\overleftarrow{\prod}}_{k,\ell}(d_{k,\ell})^{n_{k,\ell}}\right)
	|\vec{0}\kket_d
	=\delta_{\vn,\vn'}.
\end{align*}
Similarly, the biorthogonality of two odd-parity operators follows from the anti-commutation relations \eqref{eq:F-d-CAR2} and the properties of the right and left vacua $\bbra \hPi|$ and $|\vec{0}\kket_\db$ (cf.\ Prop.~\ref{prop:F-ss}).

The Liouvillian $\L=\L_+\oplus\L_-$ preserves the parity. Let us discuss the even-parity sector; the odd-parity case works in the same way.
The first term in the Jordan normal form \eqref{eq:F-Lp-Jordan} is diagonal in our basis such that
\begin{equation*}\textstyle
	\sum_{k,\ell}\xi_k d'_{k,\ell}d_{k,\ell}|\vn\kket
	=\sum_{k,\ell}\xi_k n_{k,\ell}|\vn\kket
	=\lambda_\vn|\vn\kket.
\end{equation*}
The second term yields the linear combination
\begin{equation*}\textstyle
	\sum_k\sum_{\ell=1}^{D_k-1}d'_{k,\ell}d_{k,\ell+1}|\vn\kket
	=\sum_k\sum_{\ell=1}^{D_k-1}\delta_{0,n_{k,\ell}}\delta_{1,n_{k,\ell+1}}|\tilde{\vn}_{k,\ell}\kket.
\end{equation*}
of basis states $|\tilde{\vn}_{k,\ell}\kket$ where, in comparison to $\vn$, the occupation number $n_{k,\ell+1}$ has been decreased by one, and $n_{k,\ell}$ has been increased by one. Hence, for each such term, the integer $I_\vn=\sum_{k,\ell} \ell\, n_{k,\ell}$ is decreased by one. So, the matrix \eqref{eq:F-L-upperTriag} is upper-triangular if we order the basis according to increasing $I_\vn$. This completes the proof of Prop.~\ref{prop:F-spectrum}.

Finally, let us establish the connection to the covariance-matrix evolution discussed in Sec.~\ref{sec:CovEvol}. The decay of the covariance matrix to the steady-state manifold is governed by the generator $K_-$ in Eq.~\eqref{eq:EOMvec}. For quasi-free systems, $K$ assumes the simple form
\begin{equation}
	K=X_0\otimes\id+\id\otimes X_0\quad\text{with spectrum}\quad
	\{\xi_k+\xi_{k'}\,|\,k,k'=1,\dots\NJ\}
\end{equation}
and the anti-symmetric block $K_-$ has the eigenvalues $\xi_k+\xi_{k'}$ with $k\neq k'$ if $D_k=D_{k'}=1$.
This coincides with the two-super-fermion spectrum of the full Liouvillian with $\sum n_{k,\ell}=2$ in Eq.~\eqref{eq:F-spectrum}, which makes sense as the operators $\hw_i\hw_j$ that define the covariance matrix \eqref{eq:F_G} correspond to two super-fermion excitations.
For a relaxing system, the spectral gap of the covariance-matrix evolution is hence not $\Delta$ but (bounded from below by) $2\Delta$.

\subsection{Block-triangular form for quadratic systems}\label{sec:F-blockTriangular}
While the Liouvillian spectrum \eqref{eq:F-spectrum} for a quasi-free system can be determined efficiently with a cost $\mc{O}(\Ns^3)$, this is generally not possible for quadratic systems. However, Eq.~\eqref{eq:F-L3rd} shows that all terms of their even-parity sector Liouvillians $\L_+$ do never decrease the number of super-fermions and $\L_-$ does never increase the number of super-fermions. This can be used to make the Liouvillian $\L=\L_+\oplus\L_-$ block-triangular.

Due to the anti-commutation relations \eqref{eq:F-c-CAR}, the super-fermion number operator
\begin{equation}\label{eq:F-numberOp}\textstyle
	\N_c:=\vc^\dag\cdot\vc = \sum_{j=1}^{2\Ns}c_j^\dag c_j^\pdag
\end{equation}
is self-adjoint and has all properties of a number operator with eigenvalues $N=0,\dotsc,2\Ns$. $N$ is non-decreasing and non-increasing under the action of $\L_+$ and $\L_-$, respectively. Similar to Eq.~\eqref{eq:F-basisEven}, we can construct an orthogonal basis of $\N_c$ eigen-operators \cite{Prosen2008-10}:
According to the definition \eqref{eq:F-c} of the super-ladder operators, $\bbra\id|$ is the left vacuum of all $c_j^\dag$, and the infinite-temperature state $|\vec{0}\kket_c:=\dm_\infty:=\id/2^{\Ns}$ is the right vacuum of all $c_j$. Due to $\bbra\id|\vec{0}\kket_c=\Tr( \id) / 2^{\Ns}=1$ and the anti-commutation relations \eqref{eq:F-c-CAR},
\begin{equation}\label{eq:F-basis-c}\textstyle
	|\vn\kket:={\overrightarrow{\prod}}_{j}(c^\dag_j)^{n_j}|\vec{0}\kket_c\quad\text{and}\quad
	\bbra\vn|:= \bbra\id|{\overleftarrow{\prod}}_{j}(c_j)^{n_j}\quad\text{with occupation numbers}\ \
	n_j\in\{0,1\}
\end{equation}
form a biorthogonal basis of $\N_c$ eigen-operators such that $\bbra\vn|\vn'\kket=\delta_{\vn,\vn'}$, $\N_c|\vn\kket=\sum_j n_j|\vn\kket$ and $\bbra\vn|\N_c=\sum_j n_j\bbra\vn|$, where $\vn:=(n_1,\dots,n_{2\Ns})$.

According to Eq.~\eqref{eq:F-L3rd}, the terms in $\L_+$ either keep the $\N_c$ eigenvalue $N$ invariant or increase it by two, and the terms in $\L_-$ either keep $N$ invariant or decrease it by two.
\begin{prop}[\propHead{Block-triangular form of quadratic fermionic Liouvillians}]\label{prop:F-blockTriangular}
The Liouvillian $\L=\L_+\oplus\L_-$ of quadratic fermionic systems as characterized in Eqs.~\eqref{eq:Lindblad} and \eqref{eq:F-L3rd}
assumes a block-triangular form when expressed in the eigenbasis \eqref{eq:F-basis-c} of the super-fermion number operator \eqref{eq:F-numberOp}. In particular, when ordering the basis according to increasing $N=\sum_j n_j$, the matrix representation $\bbra\vn|\L_+|\vn'\kket$ for the even-parity sector is lower block-triangular, and $\bbra\vn|\L_-|\vn'\kket$ for the odd-parity sector is upper block-triangular. The spectra of the blocks
\begin{equation}\textstyle
	\L_+|_N=\left[ \vc^\dag\cdot X_0\vc - 2\sum_s (\vc^\dag\cdot M_s\vc)^2 \right]_N\quad\text{and}\quad
	\L_-|_N=\left[ \vc\cdot X_0\vc^\dag - 2\sum_s (\vc\cdot M_s\vc^\dag)^2 \right]_N
\end{equation}
for the $N$ super-fermion sector on the diagonal yield the full spectrum of $\L$ \cite{Barthel2020_12}.
\end{prop}

In Ref.~\cite{Zhang2022-129}, this block-triangular form is used to establish that, without symmetry constraints beyond invariance under single-particle basis and particle-hole transformations, all gapped quadratic Liouvillians belong to the same phase. In many cases, one can also prove that a quadratic system is gapless by showing that one of the $N>1$ blocks has an eigenvalue with zero real part. Now, $|j\kket:=c_j|1,\dotsc,1\kket$ with $j=1,\dotsc,2\Ns$ are a basis for the $N=2\Ns-1$ super-fermion operators, and
\begin{equation}\textstyle
	 \bbra i|\L|j\kket
	 \stackrel{\eqref{eq:F-Lm-no}}{=}\bbra i|\vc\cdot X\vc^\dag|j\kket=X_{i,j}.
\end{equation}
So, the spectrum of $\L|_{2\Ns-1}$ agrees with the spectrum of the matrix $X$ in Eq.~\eqref{eq:F-EOM}.
Furthermore, $\bbra j_1,j_2|:=\bbra\id|c_{j_2}c_{j_1}$ and $|j_1,j_2\kket:=c^\dag_{j_1}c^\dag_{j_2}|\vec{0}\kket_c$ with $j_1<j_2$ form a bi-orthogonal basis for the $N=2$ super-fermion operators. Then,
\begin{align}\nonumber
	 \bbra i_1,i_2|\L|j_1,j_2\kket
	 \stackrel{\eqref{eq:F-Lp-no}}{=}&\textstyle
	   \bbra i_1,i_2|\left(\vc^\dag\cdot X\vc- 2\sum_s \sum_{i,j,k,\ell}(M_s)_{i,j}(M_s)_{k,\ell}c_i^\dag c_k^\dag c_\ell^\pdag c_j^\pdag \right)|j_1,j_2\kket\\\nonumber
	 =\,\,&  X_{i_1,j_1}\delta_{i_2,j_2}+X_{i_2,j_2}\delta_{i_1,j_1} -X_{i_1,j_2}\delta_{i_2,j_1}-X_{i_2,j_1}\delta_{i_1,j_2}\\
	  &\hspace{4ex}\textstyle - 4\sum_s\big[(M_s)_{i_1,j_1}(M_s)_{i_2,j_2}-(M_s)_{i_1,j_2}(M_s)_{i_2,j_1}\big].
\end{align}
These coincide with the matrix elements of the generator $K_-$ in the covariance equation of motion \eqref{eq:EOMvec}, i.e., with the matrix elements of $K$ with respect to the anti-symmetric vectors $(\vec{e}_{j_1}\otimes\vec{e}_{j_2}-\vec{e}_{j_2}\otimes\vec{e}_{j_1})/\sqrt{2}$.
So, the spectrum of $\L|_2$ agrees with the spectrum of the generator $K_-$.

\section{Liouvillians for bosonic systems and their Jordan normal form}\label{sec:B-Liouvillian}
In this section, we perform the steps taken for the fermionic systems in Sec.~\ref{sec:F-Liouvillian}, but now for bosonic systems. We intentionally keep the derivations as parallel as possible to allow for easy comparison.
Section~\ref{sec:B-superLadder} introduces bosonic ladder super-operators. In Sec.~\ref{sec:B-3rdQuantization}, we employ them to obtain a third-quantized form of the Liouvillians for quadratic boson systems. For the quasi-free case, Sec.~\ref{sec:B-Quasi-free} describes how to bring the Liouvillian into a many-body Jordan normal form. Such bosonic systems can be unstable, non-relaxing, and may have no steady states. Sec.~\ref{sec:B-Quasi-free-ss} yields a criterion for the existence of steady states and gives explicit expressions for them. In Sec.~\ref{sec:B-Quasi-free-Spectrum}, we obtain the full many-body spectrum by constructing a biorthogonal operator basis in which the quasi-free Liouvillian is triangular. Exploiting that the number of (a certain species of) super-bosons is non-decreasing under the action of the Liouvillian, Liouvillians of quadratic systems can be brought into a useful block-triangular form (Sec.~\ref{sec:B-blockTriangular}).

\subsection{Ladder super-operators}\label{sec:B-superLadder}
In Sec.~\ref{sec:B-Cov}, we wrote the Hamiltonian $\hH$ and Lindblad operators $\hL_r$ and $\hM_s$ in terms of the Majorana operators \eqref{eq:Majorna}, which correspond to position operators $\hw_{j+}=\hat{x}_j$ and momentum operators $\hw_{j-}=-\hat{p}_j$. To express the action of the Liouvillian on an operator $\hR$, it is convenient to define the super-operators
\begin{subequations}\label{eq:B-b}
\begin{align}
	b_{j\pm}(\hR):=&\textstyle\frac{1}{\sqrt{2}}\big(\hw_{j\pm}\hR+\hR\hw_{j\pm}\big)\quad\text{and}\\
	b'_{j\pm}(\hR):=&\textstyle\frac{\pm \mri}{\sqrt{2}}\big(\hw_{j\mp}\hR-\hR\hw_{j\mp}\big).
\end{align}
\end{subequations}
Notice that we write $b_{j\nu}'$ instead of $b_{j\nu}^\dag$ as $b_{j\nu}'\neq (b_{j\nu})^\dag$, where the adjoint is defined according to the Hilbert-Schmidt inner product \eqref{eq:HS-innerProd}. In fact, these super-operators obey
\begin{align}
	\bbra \hA|b_{j\pm}(\hB)\kket
	&\textstyle\nonumber
	=\frac{1}{\sqrt{2}}\Tr\left(\hA^\dag\big(\hw_{j\pm}\hB+\hB\hw_{j\pm}\big)\right)
	=\frac{1}{\sqrt{2}}\Tr\left(\big(\hw_{j\pm}\hA+\hA\hw_{j\pm}\big)^\dag\hB\right)
	=\bbra b_{j\pm}(\hA)|\hB\kket,\\
	\bbra \hA|b'_{j\pm}(\hB)\kket
	&\textstyle\nonumber
	=\frac{\pm \mri}{\sqrt{2}}\Tr\left(\hA^\dag\big(\hw_{j\mp}\hB-\hB\hw_{j\mp}\big)\right)
	=\frac{\pm \mri}{\sqrt{2}}\Tr\left(\big(\hw_{j\mp}\hA-\hA\hw_{j\mp}\big)^\dag\hB\right)
	=-\bbra b'_{j\pm}(\hA)|\hB\kket,\\ \label{eq:B-bAdj}
	\text{i.e.},\quad
	b^\dag_{j\pm}=&b^\pdag_{j\pm}\quad\text{and}\quad
	b'^{\dag}_{j\pm}=-b'^{\pdag}_{j\pm}.
\end{align}
The ladder super-operators \eqref{eq:B-b} obey the canonical commutation relations
\begin{equation}\label{eq:B-b-CCR}\textstyle
	[b_{i\mu},b'_{j\nu}]=\delta_{i,j}\delta_{\mu,\nu},\ \
	[b_{i\mu},b_{j\nu}]=0,\ \ \text{and}\ \
	[b'_{i\mu},b'_{j\nu}]=0\quad\text{for}\quad i,j=1,\dotsc,\Ns,\ \ \mu,\nu=\pm.
\end{equation}
The proof of is elementary. For example,
\begin{align*}\textstyle
	[b_{i\mu},b'_{j\nu}](\hR)
	=\frac{i\nu}{2}&\big(
	 \phantom{-}\hw_{i\mu}\hw_{j\bar{\nu}}\hR - \hw_{i\mu}\hR\hw_{j\bar{\nu}} + \hw_{j\bar{\nu}}\hR\hw_{i\mu} - \hR\hw_{j\bar{\nu}}\hw_{i\mu}\\
	 &         \,- \hw_{j\bar{\nu}}\hw_{i\mu}\hR - \hw_{j\bar{\nu}}\hR\hw_{i\mu} + \hw_{i\mu}\hR\hw_{j\bar{\nu}} + \hR\hw_{i\mu}\hw_{j\bar{\nu}} \big)
	=\delta_{i,j}\delta_{\mu,\nu}\hR,
\end{align*}
where $\bar{\nu}:=-\nu$, and the second step follows from the commutation relations \eqref{eq:B_MajornaCCR} of the Majorana operators.

The definition \eqref{eq:B-b} of the bosonic ladder super-operators is analogous to that for fermions in Sec.~\ref{sec:F-superLadder} and similar to corresponding definitions in Refs.\ \cite{Schmutz1978-30,Harbola2008-465,Prosen2010-43}. The approach is known as \emph{super-boson formalism} or \emph{third quantization} \cite{Note3}.

\subsection{Liouvillian in third quantization}\label{sec:B-3rdQuantization}
To express the Liouvillian \eqref{eq:Lindblad} in terms of the ladder super-operators \eqref{eq:B-b}, we first introduce the additional self-adjoint super-operators
\begin{subequations}\label{eq:B-W}
\begin{align}
	\W_{j\pm}(\hR) :=&\textstyle \hw_{j\pm}\hR
	 =\frac{1}{\sqrt{2}}\big(b_{j\pm}\pm \mri b'_{j\mp}\big)(\hR) \quad \text{and}\\
	\Wt_{j\pm}(\hR):=&\textstyle  \hR\hw_{j\pm}
	 =\frac{1}{\sqrt{2}}\big(b_{j\pm}\mp \mri b'_{j\mp}\big)(\hR)
\end{align}
\end{subequations}
The elementary Hamiltonian and linear Lindblad terms for the Liouvillian \eqref{eq:Lindblad} can now be expressed with $\W_{j\pm}$ and $\Wt_{j\pm}$ exactly as for fermions; we simply need to replace $i\mapsto i\mu$ and $j\mapsto j\nu$ in Eq.~\eqref{eq:F-Wterms}.
Defining the vectors $\vb:=(b_{1+},\dotsc,b_{\Ns+},b_{1-},\dotsc,b_{\Ns-})^T$, $\vb':=(b'_{1+},\dotsc,b'_{\Ns-})^T$, $\vW:=(\W_{1+},\dotsc,\W_{\Ns-})^T$, and $\vWt:=(\Wt_{1+},\dotsc,\Wt_{\Ns-})^T$, using the analog of Eq.~\eqref{eq:F-W-comm} and the relations
\begin{equation}\label{eq:B-W-vs-b}
	\Pmatrix{\vW\\ \vWt}=\frac{1}{\sqrt{2}}\Pmatrix{\id_{2\Ns}&-\tau\\ \id_{2\Ns}&\tau} \Pmatrix{\vb\\ \vb'},\quad
	\Pmatrix{\vW\\ \vWt}^T=\frac{1}{\sqrt{2}}\Pmatrix{\vb\\ \vb'}^T \Pmatrix{\id_{2\Ns}&\id_{2\Ns}\\ \tau&-\tau},
\end{equation}
the commutator with the Hamiltonian \eqref{eq:B_H} assumes the form
\begin{equation}\label{eq:B-L3rd-H}
	[\hH,\hR]
	= \Pmatrix{\vW\\ \vWt}    \cdot \Pmatrix{H&\\&-H}\Pmatrix{\vW\\ \vWt}(\hR)
	= \Pmatrix{\vb\\ \vb'}\cdot \Pmatrix{&-H\tau\\\tau H&}\Pmatrix{\vb\\ \vb'}(\hR)
	=2\vb'\cdot \tau H\vb\,(\hR).
\end{equation}
To see this, recall that $\tau=\sigma_y\otimes \id_{\Ns}$ as defined in Eq.~\eqref{eq:tau} and $H^T=H$ such that $\Tr(\tau H)=0$. Scalar products are denoted by $\vec{r}\cdot\vec{s}=\sum_j r_j s_j=\vec{r}^T\vec{s}$.
For the quadratic Hermitian Lindblad operators \eqref{eq:B_M} immediately follows
\begin{equation}\textstyle\label{eq:B-L3rd-M}
	\D_M(\hR)=-\frac{1}{2}\sum_s[\hM_s,[\hM_s,\hR]]
	= - 2\sum_s\big(\vb'\cdot \tau M_s\vb\big)^2\,(\hR).
\end{equation}
With the analog of Eq.~\eqref{eq:F-W-Lindblad}, $B^T=B^*$, and the relations \eqref{eq:B-W-vs-b}, the contribution of the linear Lindblad operators \eqref{eq:B_L-B} can be expressed as
\begin{align}\textstyle\nonumber
	\D_L(\hR)
	&\textstyle
	 =\sum_r \Big(\hL_r^\pdag \hR \hL_r^\dag-\frac{1}{2} \{\hL_r^\dag \hL_r^\pdag,\hR\}\Big)
	 = \left(\vW\cdot B\,\vWt-\frac{1}{2}\big(\vW\cdot B^T\vW+\vWt\cdot B\,\vWt\big)\right) (\hR)\\\nonumber
	&= \frac{1}{2}\Pmatrix{\vW\\ \vWt} \cdot \Pmatrix{-B^* & B\\B^*&-B}\Pmatrix{\vW\\ \vWt}(\hR)
	 \stackrel{\eqref{eq:B-W-vs-b}}{=}
	   \Pmatrix{\vb\\ \vb'} \cdot \Pmatrix{ 0 & 0\\ \mri \tau B_\ti & \tau B_\tr\tau}\Pmatrix{\vb\\ \vb'}(\hR)\\
	&\textstyle\label{eq:B-L3rd-L}
	 = \left(\vb'\cdot \mri \tau B_\ti\,\vb + \vb'\cdot \tau B_\tr\tau\, \vb'\right)(\hR)
\end{align}
with $B_\tr$ and $B_\ti$ as defined in Eq.~\eqref{eq:Br-Bi}.
Combining the results in Eqs.~\eqref{eq:B-L3rd-H}-\eqref{eq:B-L3rd-L}, we have the Liouvillian in terms of the ladder super-operators \eqref{eq:B-b}.
\begin{prop}[\propHead{Quadratic bosonic Liouvillians in third quantization}]\label{prop:B-3rdQuantization}
Using the ladder super-operators \eqref{eq:B-b}, the Liouvillian $\L$ in Eq.~\eqref{eq:Lindblad} for Markovian bosonic systems with quadratic Hamiltonian \eqref{eq:B_H}, linear Lindblad operators \eqref{eq:B_L-B}, and quadratic Hermitian Lindblad operators \eqref{eq:B_M} can be expressed in the form
\begin{equation}\label{eq:B-L3rd}\textstyle
	\L=\vb'\cdot X_0\,\vb + \vb'\cdot \tau B_\tr\tau\,\vb' - 2\sum_s (\vb'\cdot \tau M_s\,\vb)^2\quad\text{with}\quad
	X_0:=-2\mri\tau H + \mri\tau B_\ti.
\end{equation}
\end{prop}

The normal-ordered form of Eq.~\eqref{eq:B-L3rd} is
\begin{equation*}
	\L = \underbrace{\vb'\cdot \big(X_0-2\sum_s (\tau M_s)^2\big)\vb + \vb'\cdot \tau B_\tr\tau\,\vb'}_{
	        =\vb'\cdot X\vb + \vb'\cdot Y\vb'}
	- 2\sum_s \!\sum_{i\alpha,j\beta,m\mu,n\nu}\hspace{-2ex}(\tau M_s)_{i\alpha,j\beta}(\tau M_s)_{m\mu,n\nu}b'_{i\alpha} b'_{m\mu} b_{j\beta} b_{n\nu},
\end{equation*}
where we have the same matrices $X$ and $Y$ as in Eq.~\eqref{eq:B-EOM}. This establishes the connection to the covariance-matrix equation of motion, discussed in Sec.~\ref{sec:B-Cov}.

\subsection{Many-body Jordan normal form for quasi-free systems}\label{sec:B-Quasi-free}
For quasi-free systems, i.e., systems comprising only linear Lindblad operators $\hL_r$ but \emph{no} quadratic ones ($\hM_s$), the Liouvillian \eqref{eq:B-L3rd} assumes the form
\begin{equation}\label{eq:B-Lfree}
	\L = \frac{1}{2}\Pmatrix{\vb\\ \vb'}\cdot \Pmatrix{&X_0^T\\X_0& 2Y}\Pmatrix{\vb\\ \vb'}-\frac{\Tr X_0}{2},
\end{equation}
where $X_0=-2\mri\tau H+\mri\tau B_\ti$ and $Y=\tau B_\tr\tau$ in accordance with Eq.~\eqref{eq:B-EOM}.

It can be brought into a many-body Jordan normal form through a canonical (Bogoliubov) transformation that leads to new ladder super-operators
\begin{equation}\label{eq:B-dV}
	\Pmatrix{\vd\\ \vd'}:=V\Pmatrix{\vb\\ \vb'}\quad\text{such that}\quad
	(V^{-1})^T \Pmatrix{&X_0^T\\X_0& 2Y} V^{-1} = \Pmatrix{&\xi^T\\\xi &},
\end{equation}
where $\xi$ is a $2\Ns\times 2\Ns$ matrix in Jordan normal form with $\NJ$ Jordan blocks with eigenvalues $\xi_k$ and dimensions $D_k$. Note that, in general, $d_{k,\ell}'\neq (d_{k,\ell})^\dag$.
For the new ladder operators to obey the canonical commutation relations, we need that
\begin{equation}\label{eq:B-d-CCR}
	J_+:= \Pmatrix{ & \id_{2\Ns}\\ -\id_{2\Ns} & }
	=   \Pmatrix{[\vd,\vd^T] & [\vd,\vd'^T]   \\ [\vd',\vd^T] & [\vd',\vd'^T] } 
	= V \Pmatrix{[\vb,\vb^T] & [\vb,\vb'^{T}] \\ [\vb',\vb^T] & [\vb',\vb'^{T}] } V^T 
	= V J_+ V^T,
\end{equation}
i.e., the transformation $V$ needs to be symplectic, leaving $J_+=\mri\sigma_y\otimes \id_{2\Ns}$ invariant. 
Such a transformation does indeed exist: Let $S$ be the similarity transformation that brings $X_0$ to Jordan normal form $\xi$,
\begin{equation}
	S^{-1}X_0 S=\xi,\quad\text{and make the ansatz}\quad
	V=\underbrace{\Pmatrix{S^{-1}& \\&S^T}}_{=:V_S}
	  \underbrace{\Pmatrix{\id_{2\Ns}& -2\Gamma\\&\id_{2\Ns}}}_{=:V_\Gamma}
\end{equation}
with $\Gamma=\Gamma^T\in\RR^{2\Ns\times 2\Ns}$. This choice for $V$ obeys Eq.~\eqref{eq:B-d-CCR} as
\begin{equation*}
	J_+\,V_S^T J_+^T=\Pmatrix{S& \\&(S^{-1})^T}=V_S^{-1}\quad\text{and}\quad
	J_+\,V_\Gamma^T J_+^T=\Pmatrix{\id_{2\Ns}& 2\Gamma\\&\id_{2\Ns}}=V_\Gamma^{-1}.
\end{equation*}
And, for the Liouvillian, we obtain
\begin{equation}\label{eq:B-L-d}
	\L+\frac{\Tr X_0}{2}
	=\frac{1}{2}\Pmatrix{\vd\\ \vd'}\cdot (V^{-1})^T \Pmatrix{&X_0^T\\X_0& 2Y} V^{-1}\Pmatrix{\vd\\ \vd'}
	=\frac{1}{2}\Pmatrix{\vd\\ \vd'}\cdot \Pmatrix{&\xi^T\\\xi & Q} \Pmatrix{\vd\\ \vd'},
\end{equation}
where, as for the fermionic case (Sec.~\ref{sec:F-Quasi-free}), $Q=2S^{-1}(X_0\Gamma+\Gamma X_0^T+Y)(S^{-1})^T$ is zero if we choose $\Gamma$ to be a solution $\Gamma_\ss$ of Eq.~\eqref{eq:Gss} with $M_s=Z_s=0$. As discussed in Sec.~\ref{sec:CovEvol}, bosonic systems need not have have a steady state, i.e., in contrast to fermionic systems, we must \emph{require} that a solution $\Gamma_\ss$ exists. For the quasi-free systems, $K=X_0\otimes\id+\id\otimes X_0$ in Eq.~\eqref{eq:EOMvec}, and a unique solution $\Gamma_\ss$ exists if $K$ is invertible, i.e., if there are no two $X_0$ eigenvalues with $\xi_k+\xi_{k'}=0$.
With $Q=0$ and $\Tr X_0=\Tr\xi=\sum_k D_k \xi_k$, we then arrive at $\L=\vd'\cdot\xi\vd$.
\begin{prop}[\propHead{Jordan normal form for quasi-free bosonic Liouvillians}]\label{prop:B-L-Jordan}
The Liouvillian $\L$ in Eq.~\eqref{eq:Lindblad} for Markovian bosonic systems with quadratic Hamiltonian \eqref{eq:B_H} and linear Lindblad operators \eqref{eq:B_L-B} can be transformed to a many-body Jordan normal form if there exists a covariance matrix $\Gamma_\ss$ that solves the continuous Lyapunov equation \cite{Parks1992-9,Simoncini2016-58,Sastry1999,Khalil2002}
\begin{subequations}
\begin{equation}\label{eq:B-Gss-free}\textstyle
	X_0\Gamma_\ss+\Gamma_\ss X_0^T = -Y\quad\text{with}\quad
	X_0=-2\mri\tau H+\mri\tau B_\ti\quad\text{and}\quad
	Y=\tau B_\tr\tau.
\end{equation}
Starting from the third-quantized form \eqref{eq:B-Lfree} and using the transformation of $X_0$ to Jordan normal form $S^{-1}X_0 S=\xi$, the Liouvillian becomes
\begin{equation}\label{eq:B-L-Jordan}
	\L=\sum^\NJ_{k=1}\left(\xi_k\sum_{\ell=1}^{D_k}d'_{k,\ell}d_{k,\ell} + \sum_{\ell=1}^{D_k-1}d'_{k,\ell}d_{k,\ell+1}\right)
	\quad\!\!\text{with}\quad\!\!
	\Pmatrix{\vd\\ \vd'}=\Pmatrix{S^{-1}& \\&S^T}\Pmatrix{\id& -2\Gamma_\ss\\&\id}\Pmatrix{\vb\\ \vb'}.
\end{equation}
Here, the $k$th Jordan block of $X_0$ has eigenvalue $\xi_k$ and dimension $D_k$, $d'_{k,\ell}$ creates a super-boson (excitation) that corresponds to the generalized $\xi_k$ eigenvector of rank $\ell$, and the ladder super-operators obey the commutation relations
\begin{equation}\label{eq:B-d-CCR2}\textstyle
	[d_{k,\ell},d'_{k',\ell'}]=\delta_{k,k'}\delta_{\ell,\ell'},\quad
	[d_{k,\ell},d_{k',\ell'}]=0,\quad\text{and}\quad
	[d'_{k,\ell},d'_{k',\ell'}]=0\quad \forall k,k',\ell,\ell'.
\end{equation}
\end{subequations}
\end{prop}

\subsection{Steady states for quasi-free systems}\label{sec:B-Quasi-free-ss}
For quasi-free systems that have a \emph{physical} covariance matrix $\Gamma_\ss$ which solves the continuous Lyapunov equation \eqref{eq:B-Gss-free}, we will use the many-body Jordan normal form \eqref{eq:B-L-Jordan} of the Liouvillian \eqref{eq:B-L3rd} to establish that the Gaussian state $\dm(\Gamma_\ss)$ with covariance matrix $\Gamma_\ss$ is a steady state. If the system has eigenvalues $\xi_k=0$, further steady states exist. 
As in Sec.~\ref{sec:F-Quasi-free-ss}, we employ a Dirac notation with super-kets $|\hB\kket:=\hB$ and super-bras $\bbra\hA|$ such that $\bbra \hA|\hB\kket=\Tr(\hA^\dag\hB)$.

Gaussian states are fully characterized by their covariance matrix \cite{Wick1950-80,Danielewicz1984-152,Negele1988}. Let the covariance matrix $\Gamma$ be diagonalized by a symplectic transformation $U$ in the sense that $U\Gamma U^T=\Psmatrix{\nu&\\&\nu}$ with $\nu=\diag(\nu_1,\dotsc,\nu_n)$, and consider the transformed set of Majorana operators $\hw'_{k\kappa}=\sum_{j\nu} U_{k\kappa,j\nu}\hw_{j\nu}$ as discussed in Sec.~\ref{sec:B-Cov}. Using that a Gaussian state $\dm$ is the exponential of an operator that is quadratic in the ladder operators, $\Tr(\dm)=1$, and $\nu_k=\Tr(\dm  \hw'_{k\pm}\hw'_{k\pm})$, the corresponding Gaussian state $\dm$ can then be written explicitly in the form
\begin{equation}\label{eq:B-GaussianState}\textstyle
	\dm=\dm(\Gamma)=\prod_k\frac{1}{\nu_k+{1}/{2}}\,\left(\frac{\nu_k-{1}/{2}}{\nu_k+{1}/{2}}\right)^{\frac{1}{2}\left(\hw'^2_{k+}+\hw'^2_{k-}-1\right)}
	\quad\text{with}\quad
	\nu_k\in[1/2,\infty)
\end{equation}
and eigenvalues $\prod_k\frac{1}{\nu_k+{1}/{2}}\,\left(\frac{\nu_k-{1}/{2}}{\nu_k+{1}/{2}}\right)^{n_k}$ for $n_k\in\NN$.

\begin{prop}[\propHead{Steady states and vacua for quasi-free bosonic Liouvillians}]\label{prop:B-ss}
Consider a quasi-free bosonic system with a physical covariance matrix $\Gamma_\ss$ that solves the continuous Lyapunov equation \eqref{eq:B-Gss-free}. The corresponding Gaussian state $\dm(\Gamma_\ss)$ is then a steady state for the Liouvillian $\L$ [Eq.~\eqref{eq:B-L3rd}] and is simultaneously the right vacuum of the annihilation super-operators $d_{k,\ell}$ defined in Eq.~\eqref{eq:B-L-Jordan}, i.e.,
\begin{equation}\label{eq:B-Quasi-free-vac-d}
	d_{k,\ell}|\vec{0}\kket_d=0\ \ \forall\ k,\ell,\quad
	\L|\vec{0}\kket_d=0,
	\quad\text{and}\quad
	\bbra \id|\vec{0}\kket_d=\Tr \dm(\Gamma_\ss)=1 \quad\text{for}\quad
	|\vec{0}\kket_d:=\dm(\Gamma_\ss).
\end{equation}
If the matrix $X_0$ in Prop.~\ref{prop:B-L-Jordan} has zero eigenvalues $\xi_k=0$, then applying the corresponding creation super-operators $d'_{k,1}$ to $|\vec{0}\kket_d$ and forming linear combinations yields further (Gaussian) steady states.
Lastly, $\bbra\id|$ is the left vacuum for the creation super-operators, i.e., $\bbra\id|d'_{k,\ell}=0$ $\forall\ k,\ell$.
\end{prop}

Note that, for the Jordan normal form discussed in Prop.~\ref{prop:B-L-Jordan}, we only needed $\Gamma_\ss$ to be \emph{a} solution of the continuous Lyapunov equation \eqref{eq:B-Gss-free}. Here, we require $\Gamma_\ss$ to be \emph{physical}, in the sense that its symplectic eigenvalues lie in the range $[{1}/{2},\infty)$: For $-1/2<\nu_k<1/2$, the corresponding Gaussian operator \eqref{eq:B-GaussianState} would not be positive semidefinite and, for $\nu_k\leq 0$, it would not be a trace-class operator.

The proof of Prop.~\ref{prop:B-ss} is similar to the fermionic case discussed in Sec.~\ref{sec:F-Quasi-free-ss}. 
To see that the Gaussian state $|\vec{0}\kket_d\equiv \dm(\Gamma_\ss)=:\dm_\ss$ is the vacuum for all annihilation super-operators $d_{k,\ell}$, consider matrix elements $\bbra\id|\mc{A}\, d_{k,\ell}|\vec{0}\kket_d$ as in Eq.~\eqref{eq:F-d-matrixElements}, where $\mc{A}$ is an arbitrary product of ladder super-operators.

For bosonic Gaussian states $\dm$ and $\{\mc{D}_i\}$ being linear combinations of ladder (super-)operators, Wick's theorem \cite{Wick1950-80,Danielewicz1984-152,Negele1988} establishes that
\begin{equation}\label{eq:B-Wick}\textstyle
	\bra\mc{D}_1\dotsb \mc{D}_{2m}\ket=\sum_{\vec{i}\in\operatorname{Pair}_m} 
	\bra \mc{D}_{i_1}\mc{D}_{i_2}\ket\dotsb \bra \mc{D}_{i_{2m-1}}\mc{D}_{i_{2m}}\ket,
\end{equation}
where $\operatorname{Pair}_m$ is the set of all partitions $\vec{i}$ of numbers $\{1,\dotsc,2m\}$ into pairs without respect of order, i.e., $\vec{i}=\{(i_1,i_2),\ldots,(i_{2m-1},i_{2m})\}$ with $i_{2k-1}<i_{2k}$ $\forall\ k$.

The single-particle correlations (two-point Green's functions) needed for the evaluation of the matrix elements $\bbra\id|\mc{A}\, d_{k,\ell}|\vec{0}\kket_d$ are
\begin{equation}\label{eq:B-d-correlMatrix}
	\bbra\id| \Pmatrix{\vd\,\vd^T & \vd\,\vd'^T \\ \vd'\,\vd^T & \vd'\,\vd'^T } |\vec{0}\kket_d
	= \, V \,\bbra\id|\Pmatrix{\vb\,\vb^T & \vb\,\vb'^{T} \\ \vb'\,\vb^T & \vb'\,\vb'^{T} }|\vec{0}\kket_d \,V^T,
\end{equation}
where $V$ denotes the Bogoliubov transformation in Eqs.~\eqref{eq:B-dV} and \eqref{eq:B-L-Jordan}. Eqs.~\eqref{eq:B-b} and \eqref{eq:B-bAdj} imply that $\bbra \id|\vb'=\vec{0}$ such that the lower two blocks of the correlation matrix are zero. Due to the commutation relations \eqref{eq:B-b-CCR}, the upper-right block is $\bbra\id|\vb\,\vb'^{T}|\vec{0}\kket_d=\id_{2n} + \bbra\id|\vb'\,\vb^T|\vec{0}\kket_d=\id_{2n}$. For the upper-left block, note that
\begin{equation*}\textstyle
	b_{i\mu} b_{j\nu}(\dm_\ss)\stackrel{\eqref{eq:B-b}}{=}\frac{1}{2}\left(\hw_{i\mu}\hw_{j\nu}\dm_\ss+\hw_{i\mu}\dm_\ss\hw_{j\nu}+\hw_{j\nu}\dm_\ss\hw_{i\mu}+\dm_\ss\hw_{j\nu}\hw_{i\mu}\right),
\end{equation*}
and, hence, $\Gamma_\ss=\frac{1}{2}\bbra \id|\vb \vb^T|\vec{0}\kket_d$ according to the definition \eqref{eq:B_G}. Combining these results, the single-particle correlation matrix \eqref{eq:B-d-correlMatrix} evaluates to
\begin{equation}\label{eq:B-d-correlMatrix2}
	\bbra\id| \Pmatrix{\vd\,\vd^T & \vd\,\vd'^T \\ \vd'\,\vd^T & \vd'\,\vd'^T } |\vec{0}\kket_d
	= V \,\Pmatrix{2\Gamma_\ss & \id_{2n} \\ 0 & 0 } \,V^T 
	\stackrel{\eqref{eq:B-L-Jordan}}{=} \Pmatrix{0 & \id_{2n} \\ 0 & 0 }.
\end{equation}
Now, when expanding matrix elements $\bbra\id|\mc{A}\, d_{k,\ell}|\vec{0}\kket_d$ through Wick's theorem \eqref{eq:B-Wick} into a product of single-particle correlation functions, there is always a factor $\bra \D_i\, d_{k,\ell}\ket\equiv \bbra\id|\D_i\, d_{k,\ell}|\vec{0}\kket_d$. This factor is zero because, according to Eq.~\eqref{eq:B-d-correlMatrix2}, both $\bbra\id|\vd'\,\vd^T|\vec{0}\kket_d=0$ and $\bbra\id|\vd\,\vd^T|\vec{0}\kket_d=0$. Thus, all such matrix elements are zero, which establishes that $d_{k,\ell}|\vec{0}\kket_d=0$ $\forall\ k,\ell$.
Finally, the many-body Jordan normal form \eqref{eq:B-L-Jordan} shows that the vacuum $|\vec{0}\kket_d$ is a steady state, i.e., $\L|\vec{0}\kket_d=0$.
If $X_0$ in Prop.~\ref{prop:B-L-Jordan} has zero eigenvalues $\xi_k=0$, then further steady states are obtained by acting with the corresponding creation super-operators on $|\vec{0}\kket_d$ and forming linear combinations. In particular, $\L d'_{k,1}|\vec{0}\kket_d=d'_{k,1}\L|\vec{0}\kket_d=0$ if $\xi_k=0$.

Additionally, $\bbra\id|$ is the left vacuum of $\vd'$ as
\begin{equation}\label{B:left-d-vacuum}
	\bbra \id|\vb'\stackrel{\eqref{eq:B-b}}{=}\vec{0}\quad\Rightarrow\quad
	\bbra\id|\vd'\stackrel{\eqref{eq:B-L-Jordan}}{=}S^T\bbra\id|\vb'=\vec{0}.
\end{equation}

\subsection{Triangular form and spectrum for quasi-free systems}\label{sec:B-Quasi-free-Spectrum}
As discussed in Secs.~\ref{sec:CovEvol}, \ref{sec:F-Quasi-free-ss}, \ref{sec:B-Quasi-free}, and \ref{sec:B-Quasi-free-ss}, finite fermionic systems always have at least one steady state, whereas, for bosons, a necessary and sufficient condition is that the continuous Lyapunov equation \eqref{eq:B-Gss-free} for the steady-state covariance matrix has at least one physical solution $\Gamma_\ss$. This is assumed in the following.

\begin{prop}[\propHead{Triangular form and spectrum for quasi-free bosonic Liouvillians}]\label{prop:B-spectrum}
With the ladder super-operators for the Jordan normal form \eqref{eq:B-L-Jordan} of quasi-free bosonic Liouvillians and their corresponding left and right vacua described in Prop.~\ref{prop:B-ss}, the right and left operator bases
\begin{subequations}
\begin{equation}\label{eq:B-basis}\textstyle
	|\vn\kket:={{\prod}}_{k,\ell}\frac{1}{\sqrt{n_{k,\ell}!}}(d'_{k,\ell})^{n_{k,\ell}}|\vec{0}\kket_d,\quad
	\bbra\vn|:= \bbra\id|{{\prod}}_{k,\ell}\frac{1}{\sqrt{n_{k,\ell}!}}(d_{k,\ell})^{n_{k,\ell}}
\end{equation}
are biorthogonal, i.e., $\bbra\vn|\vn'\kket=\delta_{\vn,\vn'}$ $\forall$ $\vn,\vn'$.
Here, $\vn:=(n_{1,1},\dotsc,n_{1,D_1},n_{2,1},\dotsc,n_{\NJ,D_\NJ})$ is the vector of (super-boson) occupation numbers with $n_{k,\ell}\in\NN$. If we order the bases $\{|\vn\kket\}$ and $\{\bbra\vn|\}$ according to increasing $I_\vn:=\sum^\NJ_{k=1}\sum_{\ell=1}^{D_k} \ell\, n_{k,\ell}$, then the matrix representation
\begin{equation}\label{eq:B-L-upperTriag}
	\L_{\vn,\vn'}:=\bbra\vn|\L|\vn'\kket
\end{equation}
of the Liouvillian is upper-triangular with its diagonal containing the (generalized) eigenvalues
\begin{equation}\label{eq:B-spectrum}\textstyle
	\{\lambda_\vn = \sum^\NJ_{k=1}\sum_{\ell=1}^{D_k}\xi_k n_{k,\ell}\,|\,n_{k,\ell}\in\NN\}.
\end{equation}
The system is stable if $\Re\xi_k\leq 0$ for all generalized eigenvalues of $X_0$ in Eq.~\eqref{eq:B-Gss-free}. The system is relaxing with unique steady state $|\vec{0}\kket_d$ if $\Re\xi_k<0$ $\forall k$. The dissipative gap is then $\Delta\equiv-\max_{\vn\neq \vec{0}}\Re\lambda_\vn=-\max_k\Re\xi_k$, corresponding to a single super-boson excitation.
\end{subequations}
\end{prop}

The biorthogonality of the basis operators, denoted in the Dirac notation for the Hilbert-Schmidt inner product \eqref{eq:HS-innerProd}, follows from the commutation relations \eqref{eq:B-d-CCR}, $\vd|\vec{0}\kket_d=\vec{0}$, $\bbra \id|\vd'=\vec{0}$, and $\bbra \id|\vec{0}\kket_d=1$ (cf.\ Prop.~\ref{prop:B-ss}).
The first term in the Jordan normal form \eqref{eq:B-L-Jordan} is diagonal in our basis such that
\begin{equation*}\textstyle
	\sum_{k,\ell}\xi_k d'_{k,\ell}d_{k,\ell}|\vn\kket
	=\sum_{k,\ell}\xi_k n_{k,\ell}|\vn\kket
	=\lambda_\vn|\vn\kket.
\end{equation*}
The second term yields the linear combination
\begin{equation*}\textstyle
	\sum_k\sum_{\ell=1}^{D_k-1}d'_{k,\ell}d_{k,\ell+1}|\vn\kket
	=\sum_k\sum_{\ell=1}^{D_k-1}\sqrt{(n_{k,\ell}+1)n_{k,\ell+1}}\,|\tilde{\vn}_{k,\ell}\kket.
\end{equation*}
of basis states $|\tilde{\vn}_{k,\ell}\kket$ where, in comparison to $\vn$, the occupation number $n_{k,\ell+1}$ has been decreased by one, and $n_{k,\ell}$ has been increased by one. Hence, the matrix \eqref{eq:B-L-upperTriag} is upper-triangular if we order the basis according to increasing $I_\vn=\sum_{k,\ell} \ell\, n_{k,\ell}$.

The decay of the covariance matrix to the steady-state manifold is governed by the generator $K_+$ in Eq.~\eqref{eq:EOMvec}. Similarly to the fermionic case (Sec.~\ref{sec:F-Quasi-free-Spectrum}), for quasi-free systems, $K_+$ has the spectrum $\{\xi_k+\xi_{k'}\,|\,k,k'=1,\dots\NJ\}$. This coincides with the two-super-boson spectrum of the full Liouvillian with $\sum n_{k,\ell}=2$ in Eq.~\eqref{eq:B-spectrum}.
For a relaxing system, the spectral gap of the covariance-matrix evolution is hence not $\Delta$ but $2\Delta$, corresponding to two super-boson excitations.

\subsection{Block-triangular form for quadratic systems}\label{sec:B-blockTriangular}
For quadratic systems, there is generally no efficient way to determine the entire Liouvillian spectrum. We can however bring it into a useful block-triangular form. It is tempting to follow the steps of the fermionic case (Sec.~\ref{sec:F-blockTriangular}) by defining a super-boson number operator $\N_b=\vb'\cdot\vb$ and then work in a basis of its eigen-operators. However, $\N_b$ is not self-adjoint and, more importantly, the right vacuum of $\vb$ is the parity operator $\hPi=(-1)^{\sum_j\ha^\dag_j\ha^\pdag_j}$ which is not trace-class. It is not clear how to then construct a biorthogonal basis of $\N_b$ eigen operators.

The issue can be resolved by employing an alternative set of bosonic ladder super-operators introduced in Ref.~\cite{Prosen2010-43},
\begin{subequations}\label{eq:B-a}
\begin{alignat}{4}
	a_{j+}(\hR)&:=\ha_j \hR,\quad
	a'_{j+}(\hR)&&:=\ha_j\hR-\hR\ha_j,\\
	a_{j-}(\hR)&:= \hR\ha_j^\dag,\quad
	a'_{j-}(\hR)&&:=\hR\ha_j-\ha_j\hR.
\end{alignat}
\end{subequations}
It obeys the canonical commutation relations 
\begin{equation}\label{eq:B-a-CCR}\textstyle
	[a_{i\mu},a'_{j\nu}]=\delta_{i,j}\delta_{\mu,\nu},\ \
	[a_{i\mu},a_{j\nu}]=0,\ \ \text{and}\ \
	[a'_{i\mu},a'_{j\nu}]=0\quad\text{for}\quad i,j=1,\dotsc,\Ns,\ \ \mu,\nu=\pm.
\end{equation}
Defining the vectors $\va:=(a_{1+},\dotsc,a_{\Ns+},a_{1-},\dotsc,a_{\Ns-})^T$ and $\va':=(a'_{1+},\dotsc,a'_{\Ns-})^T$, the canonical transformation from ladder super-operators \eqref{eq:B-a} to those of Eq.~\eqref{eq:B-b} reads
\begin{equation}\label{eq:B-b-vs-a}
	\Pmatrix{\vb\\ \vb'}=\Pmatrix{\sqrt{2}\,U & U^*/\sqrt{2}\\ & U^*/\sqrt{2}} \Pmatrix{\va\\ \va'},\quad\text{where}\quad
	U:=\frac{1}{\sqrt{2}}\Pmatrix{\id_\Ns & \id_\Ns\\ \mri\id_\Ns & -\mri\id_\Ns}
\end{equation}
is a unitary $2\Ns\times2\Ns$ matrix.

The identity $\bbra\id|$ is the left vacuum of all $a'_{j\pm}$, and the zero-particle state $|\vec{0}\kket_a:=|\vec{0}\ket\bra\vec{0}|$ is the right vacuum of all $a_{j\pm}$, where $\ha_j|\vec{0}\ket=0$ $\forall j$. Due to $\bbra\id|\vec{0}\kket_a=\Tr|\vec{0}\ket\bra\vec{0}|=1$ and the commutation relations \eqref{eq:B-a-CCR},
\begin{equation}\label{eq:B-basis-a}\textstyle
	|\vn\kket:=\prod_{j\nu}(a'_{j\nu})^{n_{j\nu}}|\vec{0}\kket_a\ \ \text{and}\ \
	\bbra\vn|:= \bbra\id|\prod_{j\nu}(a_{j\nu})^{n_{j\nu}}\ \ \text{with occupation numbers}\ \
	n_{j\nu}\in\NN
\end{equation}
form a biorthogonal operator basis, similar to Eq.~\eqref{eq:B-basis}, such that $\bbra\vn|\vn'\kket=\delta_{\vn,\vn'}$.
Conveniently, they are right and left eigen-operators of the super-boson number operator
\begin{equation}\label{eq:B-numberOp-a}\textstyle
	\N_a:=\va'\cdot\va =\sum_{j=1}^{\Ns}\big(a_{j+}' a_{j+} + a_{j-}' a_{j-}\big)
\end{equation}
with $\N_a|\vn\kket=\sum_{j\nu} n_{j\nu}|\vn\kket$ and $\bbra\vn|\N_a=\sum_{j\nu} n_{j\nu}\bbra\vn|$.

Using the transformation \eqref{eq:B-b-vs-a}, we can express the Liouvillian of quadratic systems \eqref{eq:B-L3rd} in terms of the new ladder super-operators \eqref{eq:B-a}, obtaining the somewhat less compact form
\begin{align}\nonumber
	\L=&\textstyle\va'\cdot\left(\frac{1}{2}U^\dag X_0 U^*\,\va'+U^\dag X_0 U\,\va\right)
	     + \frac{1}{2}\va'\cdot U^\dag\tau B_\tr\tau U^*\,\va'\\\label{eq:B-L3rd-a}
	   &\textstyle- 2\sum_s \left[\va'\cdot\left(\frac{1}{2}U^\dag \tau M_s U^*\,\va'+U^\dag \tau M_s U\,\va\right)\right]^2.
\end{align}
The $\N_a$ eigenvalue $N$ is non-decreasing under the action of $\L$.
\begin{prop}[\propHead{Block-triangular form of quadratic bosonic Liouvillians}]\label{prop:B-blockTriangular}
The Liouvillian $\L$ of quadratic bosonic systems as characterized in Eqs.~\eqref{eq:Lindblad} and \eqref{eq:B-L3rd}
assumes a block-triangular form when expressed in the eigenbasis \eqref{eq:B-basis-a} of the super-boson number operator \eqref{eq:B-numberOp-a}. In particular, when ordering the basis according to increasing $N=\sum_{j\nu} n_{j\nu}$, the matrix representation $\bbra\vn|\L|\vn'\kket$ is lower block-triangular. The spectra of the blocks
\begin{equation}\label{eq:B-L_N}\textstyle
	\L|_N=\left[ \va'\cdot U^\dag X_0 U\,\va - 2\sum_s \left(\va'\cdot U^\dag \tau M_s U\,\va\right)^2 \right]_N
\end{equation}
for the $N$ super-boson sector on the diagonal yields the full spectrum of $\L$ \cite{Barthel2020_12}.
\end{prop}

In Ref.~\cite{Zhang2022-129}, this block-triangular form is used to establish that, without symmetry constraints beyond invariance under single-particle basis transformations, all gapped quadratic Liouvillians belong to the same phase. If one finds an $N>1$ block which has an eigenvalue with zero real part, it implies that the considered model is gapless. Now $|j\nu\kket:=a'_{j\nu}|\vec{0}\kket_a$ with $j=1,\dotsc,\Ns$ and $\nu=\pm 1$ are a basis for the $N=1$ super-boson operators and, using the normal-ordered form of Eq.~\eqref{eq:B-L_N}, one finds
\begin{equation}\textstyle
	 \bbra i\mu|\L|j\nu\kket
	 =\bbra i\mu|\va'\cdot U^\dag X U\va|j\nu\kket=(U^\dag X U)_{i\mu,j\nu}.
\end{equation}
So, the spectrum of $\L|_1$ agrees with the spectrum of the matrix $X$ in Eq.~\eqref{eq:B-EOM}.
Furthermore, one can show that the matrix elements of $\L|_2$ agree with those of $(U\otimes U)^\dag K_+ (U\otimes U)$, where $K_+$ is the generator in the covariance equation of motion \eqref{eq:EOMvec}. So, the spectrum of $\L|_2$ agrees with the spectrum of $K_+$.

\section{Applicability of Wick's theorem and closed hierarchy of correlations}\label{sec:Wick}
\subsection{Gaussian states and Wick's theorem for quasi-free systems}\label{sec:Gaussian}
Gaussian states like \eqref{eq:F-GaussianState} and \eqref{eq:B-GaussianState} are density matrices $\dm$ that can be written as the exponential of an operator that is quadratic in ladder operators \cite{Negele1988} like the Hamiltonians \eqref{eq:F_H} and \eqref{eq:B_H} or the Liouvillians \eqref{eq:F-L3rd} and \eqref{eq:B-L3rd} for quasi-free open systems (quadratic Hamiltonians $\hH$ and linear Lindblad operators $\hL_r$ only). Gaussian states can, for example, be obtained by imaginary-time or real-time evolution with a quadratic generator, starting from another Gaussian state. This corresponds to a multiplication with the exponential of a quadratic operator. As the commutator of two operators that are quadratic in ladder operators is again a quadratic operator [cf.\ Eqs.~\eqref{eq:F-comm-ww-w} and \eqref{eq:B-comm-ww-w}], the Baker-Campbell-Hausdorff formula \cite{Campbell1897-29,Baker1905-s2,Hausdorff1906,Varadarajan1984} implies that any product of such exponentials can be written as a single exponential of a quadratic operator. Hence, Gaussian states remain Gaussian under such time evolutions.
Wick's theorem shows that a Gaussian state (with vanishing first moments) is fully characterized by its covariance matrix $\Gamma$ (the two-point Green's functions) and that all $(k>2)$-point Green's functions can be computed from $\Gamma$; see Eqs.~\eqref{eq:F-Wick} and \eqref{eq:B-Wick} and Refs.~\cite{Wick1950-80,Danielewicz1984-152,Negele1988}.

According to the previous remarks, fermionic Gaussian states stay Gaussian when evolved with respect to a quasi-free Liouvillian. The steady states \eqref{eq:F-Quasi-free-vac-d} are the vacuum of a complete set of annihilation super-operators and also Gaussian: A simple explanation is that they can be obtained through an infinite time-evolution under the quadratic Liouvillian, or as the zero-temperature state of a corresponding Hamiltonian. Similarly, the excitations $|\vn\kket$ in Prop.~\ref{prop:F-spectrum} are Gaussian, because they are the vacuum for a complete set of fermionic ladder operators composed of all $d_{k,\ell}$ for $n_{k,\ell}=0$ and all $d'_{k\ell}$ for $n_{k,\ell}=1$.

As in the fermionic case, bosonic Gaussian states remain Gaussian when being evolved with a quasi-free Liouvillian, and the steady states \eqref{eq:B-Quasi-free-vac-d} are Gaussian. However, the excitations discussed in Prop.~\ref{prop:B-spectrum} are in general not Gaussian, as they do \emph{not} correspond to the vacuum of a suitable set of annihilation operators.

\subsection{Closed hierarchy of correlations for quadratic systems}\label{sec:Closedness}
For quadratic systems, i.e., open systems where the Liouvillian \eqref{eq:Lindblad} includes quadratic Hermitian Lindblad operators $\hM_s$, we have seen in Sec.~\ref{sec:Cov} that the equation of motion for the covariance matrix $\Gamma$ is still closed. That is, the time derivative of $\Gamma$ just depends on $\Gamma$ and not on $(k>2)$-point Green's functions. Nevertheless, a Gaussian initial state will generally evolve into a non-Gaussian state. For fermionic systems, this was already described in Ref.~\cite{Horstmann2013-87}. Beyond that, it turns out that the time derivative for the $k$-point Green's function tensor $\Gamma^{(k)}$ only depends on $\Gamma^{(k)}$, $\Gamma^{(k-2)}$, $\Gamma^{(k-4)}$, etc. Hence, the hierarchy of equations of motion closes and can be solved efficiently with a cost that is polynomial instead of exponential in the system size.

Let us first consider the bosonic case and, in extension of the covariance matrix \eqref{eq:B_G}, define the completely symmetric $k$-point Green's function tensors
\begin{equation}
	\Gamma^{(k)}_{j_1,\dotsc,j_k}:=\Tr\big(b_{j_1}b_{j_2}\dotsb b_{j_k}(\dm)\big)
	\equiv \bbra\id|b_{j_1}b_{j_2}\dotsb b_{j_k}|\dm\kket.
\end{equation}
The symmetry under index permutations follows from the commutation relations \eqref{eq:B-b-CCR} of the ladder super-operators \eqref{eq:B-b}. For simplicity of notation, we have dropped the Majorana indices, i.e., each index $j_n$ actually stands for a bosonic mode and a Majorana index $\pm 1$.
The time derivative is
\begin{equation}
	\partial_t \Gamma^{(k)}_{j_1,\dotsc,j_k}=\bbra\id|b_{j_1}\dotsb b_{j_k}\L|\dm\kket.
\end{equation}
The third-quantized form \eqref{eq:B-L3rd} of the Liouvillian contains terms $b'_i b_k$, $b'_i b'_k$, and $b'_i b'_k b_j b_\ell$. Due to the commutation relations \eqref{eq:B-b-CCR} and $\bbra\id|b'_i=0$ $\forall i$, the first and third Liouvillian terms couple $\partial_t \Gamma^{(k)}$ to $\Gamma^{(k)}$ and the second term couples it to $\Gamma^{(k-2)}$.

The fermionic case, restricted to the even-parity sector, can be treated in the same way. Based on the third-quantized form \eqref{eq:F-L3rd} of the Liouvillian, one finds that the time derivative of the completely anti-symmetric $k$-point Green's function tensor
\begin{equation}
	\Gamma^{(k)}_{j_1,\dotsc,j_k}:=\Tr\big(c_{j_1}c_{j_2}\dotsb c_{j_k}(\dm)\big)
	\equiv \bbra\id|c_{j_1}c_{j_2}\dotsb c_{j_k}|\dm\kket.
\end{equation}
only depends on $\Gamma^{(k)}$ and $\Gamma^{(k-2)}$, where $k$ is even.
When we allow the density matrix $\dm$ to have even and odd parity (difference) components, things become slightly more intricate. This case is covered in Ref.~\cite{Zunkovic2014-16} which provides a rather general treatment concerning the closedness of the hierarchy of equations of motion. See also Ref.~\cite{Eisler2011-06}.

\section{Stability of bosonic systems}\label{sec:B-positiveEV}
In Sec.~\ref{sec:CovEvol}, we discussed that Markovian bosonic systems need not have a steady state. The density operators still form a convex set but, even for a single mode, it is not compact. Thus, Brouwer's fixed-point theorem \cite{Brouwer1911-71,Kakutani1941-8} does not apply. 
Bosonic systems can be completely unstable, in the sense that no initial condition leads to a steady state, or partially unstable, where only some initial states decay to a steady state. Physically, unstable systems correspond to situations where the environment indefinitely pumps energy or particles into the system.
A simple example for the partially unstable scenario is a system with a Lindblad operator $\ha^\dag_1\ha^\pdag_1\ha^\dag_2$. The environment would endlessly pump particles into mode $2$, if there is at least one particle in mode $1$. But it is stable if mode $1$ is unoccupied in the initial state.

As a specific example for an unstable quasi-free system, consider a single mode, no Hamiltonian, and the linear Lindblad operator $\hL=\ha^\dag\stackrel{\eqref{eq:Majorna}}{=}(\hw_+ +\mri\hw_-)/\sqrt{2}$ being the creation operator. With the corresponding vector $\vec{L}=(1,\mri)^T/\sqrt{2}$,
we have $B=\vec{L}\vec{L}^\dag=\frac{1}{2}\Psmatrix{1&-\mri\\ \mri&1}$ in Eq.~\eqref{eq:B_L-B}. The real and imaginary parts \eqref{eq:Br-Bi} of this matrix are
\begin{equation}
	B_\tr=\frac{1}{2}\Pmatrix{1&\\&1}\quad\text{and}\quad
	B_\ti=\frac{1}{2}\Pmatrix{&-1\\1&}
\end{equation}
such that the continuous Lyapunov equation \eqref{eq:B-Gss-free} reads
\begin{equation}
	X_0\Gamma_\ss+\Gamma_\ss X_0^T = -Y = -\tau B_\tr\tau = -\id /2 \quad\text{with}\quad
	X_0=\mri\tau B_\ti=\id/2
\end{equation}
as, for a single mode, $\tau=\sigma_y$. The unique solution is $\Gamma_\ss=-\id/2$. With this solution, we can bring the Liouvillian into Jordan normal form according to Prop.~\ref{prop:B-L-Jordan}. However, this $\Gamma_\ss$ is \emph{not} a physical covariance matrix, as its unique symplectic eigenvalue $-1/2$ is outside the range $[{1}/{2},\infty)$. Hence, Prop.~\ref{prop:B-ss} is not applicable and there exists no steady state. Specifically, the corresponding density operator \eqref{eq:B-GaussianState} would not be trace-class. Physically, the coupling to the environment leads to an ever-increasing particle number in this system. This instability is also reflected in the fact that $X_0$ has the positive eigenvalue $\xi=1/2$ and that, consequently, the generator $K_+$ of the covariance matrix evolution \eqref{eq:EOMvec} has the positive eigenvalue $\kappa=1$. The annihilation operator is a left eigen-operator of the Liouvillian with eigenvalue $1/2$,
\begin{equation}\textstyle
	\L^\dag(\ha)=\D_L^\dag(\ha) \stackrel{\eqref{eq:LindbladAdj-DL}}{=}
	\ha \ha \ha^\dag-\frac{1}{2} \{\ha \ha^\dag,\ha\}=\frac{1}{2}\ha.
\end{equation}

The case of quasi-free systems is special in the sense that the existence of a steady state implies stability.
\begin{prop}[\propHead{Stability of quasi-free bosonic systems}]\label{prop:B-stability}
A quasi-free bosonic system is stable if its Lyapunov equation has a physical solution $\Gamma_\ss$.
\end{prop}
Recall from Sec.~\ref{sec:B-Cov} that a bosonic covariance matrix is called physical if its symplectic eigenvalues lie in the range $[{1}/{2},\infty)$. Also recall from Sec.~\ref{sec:B-Quasi-free-ss} that a quasi-free bosonic system has a steady state if the continuous Lyapunov equation \eqref{eq:B-Gss-free} has a physical solution $\Gamma_\ss$. If a steady state exists, then, according to the results on the many-body spectrum \eqref{eq:B-spectrum}, the system will be stable if all eigenvalues of the structure matrix $X_0$ have non-positive real parts. Now, assume that $X_0$ has a left eigenvector $\vec{\xi}^\dag$ with eigenvalue $\xi$. From the Lyapunov equation \eqref{eq:B-Gss-free}, we then obtain
\begin{equation}
	-\vec{\xi}^\dag Y\vec{\xi} = \vec{\xi}^\dag \big(X_0\Gamma_\ss+\Gamma_\ss X_0^T\big) \vec{\xi} = 2\Re(\xi)\,\vec{\xi}^\dag\Gamma_\ss\vec{\xi}
\end{equation}
The left-hand side is non-positive, because $Y=\tau B_\tr\tau$ is positive semidefinite according to Eq.~\eqref{eq:B_L-B}. As discussed in Sec.~\ref{sec:B-Cov}, a physical covariance matrix $\Gamma_\ss$ is positive definite such that $\vec{\xi}^\dag\Gamma_\ss\vec{\xi}> 0$ and, hence, $\Re(\xi)\leq 0$. This proves the proposition.

\section{Discussion}\label{sec:Discuss}
We have described a comprehensive framework for solving quasi-free and quadratic Lindblad master equations of Markovian fermionic and bosonic systems. The equations of motion for $k$-particle Green's functions form closed hierarchies and can hence be solved efficiently. With a formalism of ladder super-operators, the Liouvillians can be brought into a convenient third-quantized form that reveals a block-triangular structure. For the quasi-free case, canonical transformations lead to a many-body Jordan normal form and a biorthogonal operator basis in which the Liouvillian is triangular, revealing the full many-body spectrum. Fermionic systems always have at least one steady state. For bosonic systems, this is not necessarily the case, and we described criteria for the existence of steady states, the stability of the systems, and their relaxation properties.

Beyond the study of specific models, this systematic treatment may be useful for further developments of perturbative and renormalization-group techniques, exact solutions, and numerical methods. Our motivation for this work was an investigation of criticality and dissipative phase transitions in quadratic open systems. Corresponding results are presented in the companion paper \cite{Zhang2022-129}.

\begin{acknowledgments}
We gratefully acknowledge support through US Department of Energy grant DE-SC0019449.
\end{acknowledgments}

\end{document}